\documentclass[prd,preprint,tightenlines,floatfix,showpacs,preprintnumbers,nofootinbib,eqsecnum]{revtex4}

 \usepackage[dvips,final]{graphicx}
  \usepackage{amssymb}
   \usepackage{amsmath}
    \usepackage{amsfonts}
     \usepackage{epsfig}
      \usepackage{bm} % bold math

\begin{document}

\title{Light meson gas in the QCD vacuum and oscillating Universe\footnote{In memory of Prof. G.~M.~Vereshkov.}}

\author{George Prokhorov$^1$}
\email{prokhorov@theor.jinr.ru}

\author{Roman Pasechnik$^{2,3}$}
\email{Roman.Pasechnik@thep.lu.se}

\affiliation{
{$^1$\sl 
Bogoliubov Laboratory of Theoretical Physics, 
Joint Institute for Nuclear Research, Dubna, Russia
}\vspace{0.5cm}\\
{$^2$\sl
Department of Astronomy and Theoretical Physics, Lund
University, SE-223 62 Lund, Sweden
}\vspace{0.5cm}
\\
{$^3$\sl
Nuclear Physics Institute ASCR, 25068 \v{R}e\v{z}, Czech Republic
}\vspace{0.5cm}
%\footnote{Also at Nuclear Physics Institute ASCR, 25068 \v{R}e\v{z}, Czech Republic}}
}
%
%
%
%%%%%%%%%%%%%%%%%%%%%%%%%%%%%%%%%%%%%%%%%%%%%%%%%%%%
\begin{abstract}
\vspace{0.5cm}
We have developed a phenomenological effective quantum-field theoretical model describing the ``hadron gas'' of the lightest 
pseudoscalar mesons, scalar $\sigma$-meson and $\sigma$-vacuum, i.e.~the expectation value of the $\sigma$-field, 
at finite temperatures. The corresponding thermodynamic approach was formulated in terms of the generating functional 
derived from the effective Lagrangian providing the basic thermodynamic information about the ``meson plasma + QCD condensate'' 
system. This formalism enables us to study the QCD transition from the hadron phase with direct implications for cosmological 
evolution. Using the hypothesis about a positively-definite QCD vacuum contribution stochastically produced in early universe, 
we show that the universe could undergo a series of oscillations during the QCD epoch before resuming unbounded expansion.
\end{abstract}
%%%%%%%%%%%%%%%%%%%%%%%%%%%%%%%%%%%%%%%%%%%%%%%%%%%%


\maketitle

%================
\section{Introduction}
\label{Sec:Intro}
%================

The Linear Sigma Model (L$\sigma$M) \cite{GellMann:1960np} is well-known as a useful and simple approach to study QCD thermodynamics in the strongly-coupled 
regime and, in particular, the chiral symmetry restoration, see e.g.~Refs.~\cite{Pisarski:1983ms, Rischke:2003mt, Bowman:2010zz, Mocsy:2004ab, Scavenius:2000qd}. 
Already in simplest realisations based on the L$\sigma$M a strong phase transition between different phases of the strongly-interacting matter has been found to occur 
\cite{Bowman:2010zz}. Notably, it was shown that incorporation of fermions to the model by means of the bosonization approximation changes the phase 
transition to crossover \cite{Bowman:2010zz, Mocsy:2004ab} and brings the model to a conformity with the predictions obtained by lattice simulations 
\cite{Aoki:2006we}. The effective L$\sigma$M-based approaches thus prove to be suitable for description of the interacting meson plasma and can be applied 
for studies of the cosmological QCD phase transition.

The current work extends the previous study of Ref.~\cite{Vereshkov:1995yt} where the method of generating functional was applied in analysis of the electroweak phase 
transition in the framework of classical and effective non-classical Higgs models. We start with a new form of the effective L$\sigma$M-type Lagrangian describing 
the spectrum of lightest pseudoscalar mesons such as $\pi^\pm$, $\pi^0$, $K^\pm$, $K^0$, $\bar{K}^0$, $\eta$, $\eta'$ as well as the scalar $\sigma$-meson 
(with scalar $f_0(500)$ state as the closest candidate \cite{Agashe:2014kda}). In this model, the quartic self-interactions are accounted for the $\sigma$-meson only 
while the quartic terms for pseudoscalar mesons are omitted. This approximation corresponds to the ``hadron gas'' of pseudoscalar mesons interacting with the
non-linear $\sigma$-field. Below we show that such a simplified version of the L$\sigma$M is sufficient to reproduce relevant thermodynamic properties of 
the system such as $\sigma$-condensate ``melting'' and effective meson mass dynamics previously found in various L$\sigma$M-based theories, see 
e.g.~Refs.~\cite{Bowman:2010zz,Mocsy:2004ab,Chen:2013oya}, or in other theoretical approaches, for example, in the Nambu--Jona-Lasinio (NJL) and 
Polyakov-loop extended NJL (PNJL) models \cite{Blaschke:2014zsa, Buballa:2003qv}.

The gas of pseudoscalar mesons, however, exhibits a few important features that have not been previously discussed. One of them concerns the specific properties 
of the meson spectrum after chiral symmetry restoration when mesons supposedly can be considered as metastable states with a particular width 
\cite{Nahrgang:2013xaa,Turko:2014jta}. Below we show that the pseudoscalar meson masses remain finite and approximately constant (or slowly increase) 
above the critical temperature $T_c$ leaving a possibility for $\sigma\to\pi^+\pi^-$ and $\sigma\to\pi^0\pi^0$ decays.

Usually, thermodynamic information cannot be fully extracted from a particular quantum-field theoretical model as for that purpose one has to compute an infinite series 
of diagrams emerging from the partition function constructed in the form of Euclidean path integral \cite{Kapusta:2006pm}. At this stage, one could use either perturbative 
expansions or non-perturbative methods. One of non-perturbative approaches called the Carter's method \cite{Bowman:2010zz} corresponds to resummation 
of daisy and superdaisy diagrams \cite{Carter:1998ti,Bowman:2010zz,Vereshkov:1995yt} and is equivalent to the $\Phi$ Derivable Approximation. This method was 
widely used in different applications, in particular, for gluon plasma thermodynamics \cite{Carter:1998ti}, for real-time dynamics of the homogeneous condensates in 
Yang-Mills theories \cite{Prokhorov:2013xba}, for a QCD transition analysis in the L$\sigma$M \cite{Bowman:2010zz,Mocsy:2004ab} and for electroweak phase transition 
in the Higgs model \cite{Vereshkov:1995yt}. 

The Carter's method allows to go beyond the mean-field approximation and to take into account contributions of the fluctuations and their back-reaction to the mean-field 
medium. We apply the Carter's method to our model in order to investigate certain thermodynamic properties of the meson plasma such as the phase structure, effective meson 
masses, energy, pressure, fluctuation amplitudes and speed of sound.

As it was shown in Refs.~\cite{Bowman:2010zz,Mocsy:2004ab}, the microscopic evolution of the ``meson gas + condensate'' system described by equations of motion 
can be reproduced by minimization of the free-energy functional (generating functional) over all its parameters while the stability analysis of the phases is governed 
by the non-equilibrium Landau functional which can be directly constructed from the generating functional. The minima of the Landau functional correspond to 
equilibrium states of the system. These two functionals are intensively used in our analysis below.

Our approach and methodology are similar to those of Ref.~\cite{Bowman:2010zz}. However, we take into account the zero-point fluctuations, as it was done in 
Ref.~\cite{Mocsy:2004ab}. For simplicity, we do not include fermions in this work as was previously done in Refs.~\cite{Bowman:2010zz,Mocsy:2004ab}, and as an expected 
result we get the first-order phase transition instead of crossover. In comparison to Ref.~\cite{Mocsy:2004ab}, however, we consider additional mesonic degrees of freedom 
and compute various thermodynamic quantities. These and other differences in the form of our effective Lagrangian lead to a modification of thermodynamic 
properties of the considered meson plasma system that requires a proper discussion.

The constructed model is characterised by a set of the parameters which can be fixed using the experimentally measured characteristics of light meson spectrum in 
the vacuum and known amplitude of the QCD condensate. However, as the properties of $\sigma$-meson are still not unambiguosly fixed by measurements, we study 
the dependence of the results on the value of $\sigma$-meson vacuum mass. We introduce $\eta'$-meson in a somewhat different way in comparison to other 
pseudoscalars since its mass is modified by the gluon anomaly contribution. For the chosen set of physical parameters we obtain too large value of the critical 
temperature $T_c\simeq 400\, {\rm MeV}$ limiting the predicting power of the model to a qualitative level. We expect that the value of critical temperature 
will decrease after inclusion of fermions (quarks and baryons) as was noticed in Ref.~\cite{Mocsy:2004ab}, as well as other potentially relevant hadronic degrees 
of freedom. Despite such quantitative limitations of our analysis, its qualitative predictions can be important e.g. for cosmological implications of the considered 
``(pseudo)scalar meson gas + condensate'' system. Our simplified effective model enables us to study the basic features of the cosmological QCD transition 
and to extract the most relevant characteristics of the cosmological expansion at that epoch.

As is well-known, during the first second after the Big Bang the universe has experienced a series of relativistic thermodynamic transitions followed by formation of 
new vacuum subsystems and by breaking of the fundamental symmetries. One of the most well-studied transitions has happened in the QCD sector at the moment 
$t\sim 10^{-5}\,\rm{s}$ after the Big Bang when the temperature of the universe was about $T\sim 200 \,{\rm MeV}$ and the Hubble radius was 
$d_{H}\approx 10\,{\rm km}$. Up to now, the cosmological QCD phase transition was investigated in various aspects such as the order of the transition, 
the scale-parameter evolution, the generation of density fluctuations and a role of the background magnetic fields. For a thorough review on these problems, 
we refer an interested reader to e.g.~Refs.~\cite{Bonometto:1993pj,Schwarz:2003du,Boyanovsky:2006bf,Castorina:2015ava,Tawfik:2008cd} and references therein.

While the QCD phase transition was thoroughly studied in cosmology for many years, the problem of the negatively-definite QCD vacuum energy has received a little attention.
From the QCD instantons theory we know that the QCD transition is followed by formation of the negative quark-gluon vacuum denisty \cite{Shifman:1978bx,Schafer:1996wv} 
which is of primary importance for hadron physics. In particular, this condensate provides the mass generation mechanism of light mesons as it may be seen 
from the Gell-Mann--Oakes--Renner relation \cite{GellMann:1968rz,Reinders:1984sr,Ioffe:1981kw}. The fact which we would like to bring up for a discussion here is that 
in cosmology the vacuum contributions with such a big negative energy density lead to a fundamental problem \cite{Bull:2015stt,Pasechnik:2013sga,Pasechnik:2016sbh,
Pasechnik:2016twe} since the right hand side of the Friedmann equation has to be positive in the case of flat universe. This is one of the implications of the well-known 
vacuum catastrophe \cite{Weinberg:1988cp} which points to a fundamental inconsistency of cosmological observations with typical values of the vacuum energy density
predicted in the framework of microscopic quantum field theory. It can be directly shown that the presence of a negative QCD vacuum contribution immediately 
leads to a very fast collapse of the universe, such that it never passes through the cosmological horizon corresponding the QCD transition time scale. Below, such 
an effect will be shown in the framework of the considered effective meson plasma model as well as in the Bag-like model of the QCD crossover \cite{Ferroni:2008ej}. 
As the universe collapses, the temperature increases up to the critical value $T_c$ and then a ``backward'' transition to the quark-gluon plasma occurs. 
The existence of such a second ``backward'' transition is yet another important prediction stemming from the presence of the evolving (with temperature) 
QCD vacuum in the cosmological medium.

Then, one can discuss possible consequences of such two-transition cosmological evolution directly following from QCD thermodynamics. It is clear that fast collapse 
at the QCD scale would not allow the modern large-scale universe to be formed. Therefore, another compensating positively-definite vacuum contribution has to be formed 
during the QCD transition epoch and to co-exist with the negatively-definite one. For example, in Refs.~\cite{Pasechnik:2013sga} and \cite{Pasechnik:2016twe} it has been 
shown that under certain assumptions about a particular form of the non-perturbative Yang-Miils gauge coupling one finds a generic non-perturbative 
solution for non-Abelian fields corresponding to a homogeneous vacuum ($\Lambda$-term) with positive energy density. The latter could, in principle, 
compensate the usual quantum-topological gluon condensate at the QCD energy scale. Under certain rather general assumptions about a stochastic production 
mechanism of such a positively-definite vacuum contribution, we come to the concept of oscillating universe when a series of (potentially many) oscillations of 
the universe happen around the QCD transition epoch. We present a first qualitative picture of the corresponding cosmological evolution.

The structure of the paper is as follows. In Sect.~\ref{Sec:Meson gas model}, the generic properties of the QCD vacuum and meson interactions are discussed and
the effective L$\sigma$M for the interacting meson plasma and QCD condensate is formulated. In Sect.~\ref{Sec:Thermodynamics}, the thermodynamic approach 
to the meson plasma based upon the generating functional is developed. Then, in Sect.~\ref{Sec:num}, we obtain and study numerically the basic thermodynamic 
observables. In Sect.~\ref{Sec:Cosmology}, the main consequences of the presence of the QCD vacuum (evolving with temperature) in cosmology are considered 
in the framework of the interacting meson plasma model as well as the Bag-like model, and a particular scenario of the oscillating universe has been discussed. 
In Sect.~\ref{Sec:summary}, concluding remarks are given.

%%%%%%%%%%%%%%%%%%%%%%%%%%%%%%%%%%%%
\section{Effective theory of the meson gas}
\label{Sec:Meson gas model}
%%%%%%%%%%%%%%%%%%%%%%%%%%%%%%%%%%%%

%%%%%%%%%%%%%%%%%%%%%%%%%%%%%%%%%
\subsection{QCD condensate and scalar mesons in the vacuum}
\label{Sec:QCD condensate}
%%%%%%%%%%%%%%%%%%%%%%%%%%%%%%%%%

The order parameter in QCD, the quark-gluon condensate, acquires contributions from the quark and gluon quantum fluctuations 
whose typical energy density at zero temperature
\begin{eqnarray}
\label{eps-QCD}
\epsilon_{\rm top}(T=0) \equiv \epsilon^{\rm gluon}+\epsilon^{\rm quark} \simeq -(5\pm 1)\times 10^{9}\;\text{MeV}^4 \,,
\end{eqnarray} 
where the dominating gluon contribution spontaneously appears as a result of quantum tunnelling processes between various 
topological states of the gluon field and corresponds to the trace anomaly in the gluonic energy-momentum tensor
\begin{eqnarray}
\label{eps-gluon}
\epsilon^{\rm gluon} \equiv \frac{1}{4}\,\langle 0| T^{\mu,{\rm gluon}}_{\mu}  |0 \rangle \,, \qquad 
T_{\mu}^{\mu,{\rm g}}=\frac{\beta({\bar g}_s^2)}{2}\,G_{\mu\nu}^aG^{\mu\nu}_a \,,
\end{eqnarray} 
defined as an average over the quantum state vector $|0 \rangle$ in terms of the QCD $\beta$-function $\beta=\beta({\bar g}_s^2)<0$, 
gluonic energy-momentum $T^{\nu,{\rm gluon}}_{\mu}$ and stress $G^{\mu\nu}_a$ tensors (for $SU(N_c)$ gauge theory $a=1\dots N_c^2-1$), 
and the QCD running coupling ${\bar g}_s^2={\bar g}_s^2(\mu^2)$. In Eq.~(\ref{eps-gluon}), the subleading quark contribution to 
the QCD ground state density reads
\begin{eqnarray}
\label{eps-quark}
&&\epsilon^{\rm quark}=\frac{1}{4}\langle 0|m_u \bar{u}u + m_d\bar{d}d + m_s\bar{s}s|0\rangle \,, \\
&& \langle 0 |\bar{s}s|0\rangle\simeq \langle 0 |\bar{u}u|0\rangle = \langle 0 |\bar{d}d|0\rangle =
- l_g\langle 0 |\frac{\alpha_s}{\pi}G^{a}_{\mu\nu}G_{a}^{\mu\nu}|0\rangle = -(235\pm 15\, \rm{MeV})^3 \,,
\label{q-vs-g}
\end{eqnarray}
and corresponds to the net effect of the light sea quark $u,d,s$ quantum fluctuations induced by the gluon fluctuations at typical length 
scale $l_g$ known as the gluon correlation length. To one-loop order \cite{Shifman:1978bx,Schafer:1996wv}, one writes
\begin{eqnarray}
\label{eps-gluon-1L}
\beta({\bar g}_s^2)=-\frac{b{\bar g}_s^2}{16\pi^2}\,, \quad {\bar g}_s^2=\frac{16\pi^2}{\displaystyle b\ln(\mu^2/\Lambda_{\rm QCD}^2)}\,, 
\quad \epsilon^{\rm gluon} = -\frac{b}{32}\langle0|\frac{\alpha_s}{\pi}G^a_{\mu\nu}G_a^{\mu\nu}|0\rangle \,,
\end{eqnarray} 
where $\alpha_s={\bar g}_s^2/4\pi$, $\Lambda_{\rm QCD}\simeq 0.28$ GeV is the QCD scale parameter, and the one-loop coefficient of 
the $\beta$-function in QCD is $b=9$. The characteristic momentum scales $\mu\sim \Lambda_g$ inverse to the gluon correlation length $l_g$, 
i.e. $\Lambda_g \sim l_g^{-1}\simeq 1.2$ GeV \cite{Shifman:1978bx,Schafer:1996wv}, where the perturbative QCD still provides a realistic estimate, 
determine the formation and dynamics of the gluon condensate $\langle G^2 \rangle>0$ and, hence, the key properties of the hadronic medium.

The topological quark and gluon field fluctuations, their number densities and interactions between them depend on the state of hadronic matter 
in which they appear. In the vacuum (i.e. in the absence of particles at $T=0$) the fluctuations are characterised by the net QCD condensates' density
\begin{eqnarray}
\epsilon_{\rm top}(T=0)=-\Big(\frac{9}{32}+\frac{(m_u+m_d+m_s)l_g}{4}\Big)
\langle 0 |\frac{\alpha_s}{\pi}G^{a}_{\mu\nu}G_{a}^{\mu\nu}|0\rangle\equiv -v_0^4 \,,
\label{eps-QCD-vac}
\end{eqnarray}
where the value of the QCD condensates' amplitude $v_0$ estimated as
\begin{eqnarray}
\label{v0}
v_0=265\pm 15\, \rm{MeV} \,.
\end{eqnarray}
In the hadron plasma at finite temperature $T$, the QCD condensates' density should be different since real hadrons can influence the tunnelling 
processes in the ground state. Then the parameter $v=v(T)$ defining the QCD condensates' density in the hadron plasma
\begin{eqnarray}
\epsilon_{\rm top}(T) = -v(T)^4 \,, \qquad v(0)=v_0 \,,
\label{eps-QCD-vac-T}
\end{eqnarray}
can be considered as the QCD order parameter for the hadron matter.

The mechanism of generation of the lightest meson mass spectrum is related to the chiral symmetry breaking in QCD. Formally, three light $u,d,s$ quark system 
is invariant under the global chiral $U_{\rm L}(1)\times SU_{\rm L}(3)\times U_{\rm R}(1)\times SU_{\rm R}(3)$ symmetry. A part of this symmetry 
$SU_{\rm L}(3)\times SU_{\rm R}(3)$ is broken both explicitly (by the small $u,d,s$ current quark masses) and spontaneously (by the light sea quark 
condensates (\ref{q-vs-g}) generated by the spontaneously-induced gluon condensate). The lightest pseudoscalar $\pi,\,K$ and $\eta$ mesons are considered 
as pseudo-Goldstone bosons whose masses are determined by the quark condensates. According to the well-known Gell-Mann-Oaks-Renner relation \cite{GellMann:1968rz}, 
the meson masses at zero temperature are expressed in terms of the light quark condensates and the meson decay constants as follows
\begin{eqnarray}
&& m_{\pi \rm{(vac)}}^2=-\frac{(m_u+m_d)\langle 0|\bar{u}u+\bar{d}d|0\rangle}{f^2_{\pi}}\,, \nonumber \\
&& m_{K \rm{(vac)}}^2=-\frac{(m_u+m_d+2m_s)\langle 0|2\bar{s}s+\bar{u}u+\bar{d}d|0\rangle}{4f^2_{K}}\,,  \label{masses}\\
&& m_{\eta \rm{(vac)}}^2=-\frac{(m_u+m_d+4m_s)\langle 0|4\bar{s}s+\bar{u}u+\bar{d}d|0\rangle}{9f^2_{\eta}}\,, \nonumber
\end{eqnarray}
for $\pi,\,K,\,\eta$ mesons, respectively. In the approximate calculations below, we use the following reasonable 
relation for meson decay constants
\begin{eqnarray}
f_{\pi}^2 \simeq f_{K}^2\simeq f_{\eta}^2\simeq (130\, \rm{MeV})^2\,.
\label{dec-const}
\end{eqnarray}
which together with Eq.~(\ref{q-vs-g}) is justified by the approximate (within 6.5\%) phenomenological 
relation between the masses
\begin{eqnarray}
4m_K^2\simeq 3m_\eta^2+m_\pi^2 \,.
\end{eqnarray}

For the $\eta'$ meson the situation is different. If another part of the chiral symmetry $U_{\rm L}(1)\times U_{\rm R}(1)$ would be explicitly 
broken only by the small current quark masses in the same way as $SU_{\rm L}(3)\times SU_{\rm R}(3)$, then the $\eta'$ meson would also be 
a pseudo-Goldstone boson and its mass would be related to those of $K$ and $\pi$ mesons as 
\[ m_{\eta'}\simeq \frac{1}{3}(2m_K^2+m_\pi^2) \]
which, however, strongly contradicts to the experimental data. This means that $U_{\rm L}(1)\times U_{\rm R}(1)$ is broken
much stronger than it would be broken by the presence of the current quark masses only. The strong explicit breaking of $U_{\rm L}(1)\times U_{\rm R}(1)$ 
symmetry is known to be induced by the gluon trace anomaly which is an integral characteristics of the topological gluon structures participating in tunnelling processes. 
In order account for this effect, an additional length parameter $l_{\rm an}\equiv \Lambda_{\rm an}^{-1}$ with the energy scale of the gluon anomaly 
$\Lambda_{\rm an}\simeq 0.5\,\rm{GeV}$ characterising its contribution to the $\eta'$ mass can be introduced such that
\begin{eqnarray}
m_{\eta' \rm{(vac)}}^2=-\frac{4(m_u+m_d+m_s+\Lambda_{an})\langle 0|\bar{s}s+\bar{u}u+\bar{d}d|0\rangle}{9f^2_{\eta'}}\,, \label{mass-eta-p}
\end{eqnarray}
The length scale $l_{\rm an}$ is a typical size of topological configurations participating in tunnelling processes while the correlation length $l_g$
is the characteristic space-time scale of the gluon field fluctuations formed during the tunnelling. The pseudoscalar $\pi,\,K,\,\eta$ and $\eta'$ mesons
are considered as various quantum correlation waves between the quark and antiquark fluctuations of the QCD condensate. Note, Eqs.~(\ref{masses}) and 
(\ref{mass-eta-p}) provide the vacuum values of the masses of the pseudoscalar mesons in the hadronic medium at zero temperature $T=0$. 

It is quite natural to assume linear dependence of the meson decay constants on the order parameter at $T=0$ (\ref{v0}). Indeed, $f_\pi\simeq 130$ MeV is directly
related to the breaking of the chiral $SU_{\rm L}(3)\times SU_{\rm R}(3)$ symmetry and, hence, to the corresponding density of the topological quark-gluon
fluctuations. Thus, defining $f_{\pi}\equiv v_0/k_{\pi}$ with $k_{\pi}\simeq 2$, one can rewrite the meson mass relations as follows
\begin{eqnarray}
&& m_{\pi \rm{(vac)}}^2=2\kappa(m_u+m_d)v_0^2\,, \quad 
m_{K \rm{(vac)}}^2=\kappa(m_u+m_d+2m_s)v_0^2\,,  \nonumber \\
&& m_{\eta \rm{(vac)}}^2=\frac{2}{3}\kappa(m_u+m_d+4m_s)v_0^2\,, \quad 
m_{\eta' \rm{(vac)}}^2=\frac{4}{3}\kappa(m_u+m_d+m_s+\Lambda_{\rm an})v_0^2\,, 
\label{masses-v0}
\end{eqnarray}
where
\begin{eqnarray}
\kappa \equiv 4 k_{\pi}^2\Big(\frac{9}{8}\Lambda_g+m_u+m_d+m_s\Big)^{-1}\,,
\label{kappa}
\end{eqnarray}
which will be useful in what follows.

%%%%%%%%%%%%%%%%%%%%%%%%%%%%%%%%%%%%%%%%%%%
\subsection{Effective Lagrangian, equations of motion and meson mass spectrum}
\label{Sec:Lagrangian}
%%%%%%%%%%%%%%%%%%%%%%%%%%%%%%%%%%%%%%%%%%%

In the previous subsection we have discussed the essential properties of low-energy QCD and established a connection between
the QCD order parameter at $T=0$, $v_0$, and the pion decay constant, $f_\pi$, leading to quadratic dependence between 
the meson masses squared and $v_0$.

In the L$\sigma$M the QCD condensate can be described by the self-interacting $\sigma$-field with non-zeroth VEV 
$\langle \sigma \rangle\not=0$ defined as an average of the $\sigma$-field over the quantum state vector $|0 \rangle$ 
(a shorthand notation $\langle\, \dots \rangle\equiv \langle 0| \dots |0 \rangle$ is used here and below). Following the Carter's 
method \cite{Carter:1998ti}, ones decomposes the $\sigma$-field into the condensate ($\sigma$-VEV) and wave fluctuations
about the condensate, $\tilde{\sigma}\equiv \sigma - \langle \sigma \rangle$. The Lagrangian of the $\sigma$-field having 
the quartic Higgs-like potential reads
\begin{eqnarray}
\label{Lsig}
\mathcal{L}_\sigma = \frac{1}{2}\partial_{\mu}\sigma\partial^{\mu}\sigma - U(\sigma) \,, \qquad
U(\sigma)=-2g^2 v_0^2 \sigma^2+g^4 \sigma^4 \,,
\end{eqnarray}
the $g$ is the $\sigma$ quartic coupling. Introducing the $\sigma$-VEV as
\begin{eqnarray}
\sigma = \langle \sigma \rangle + \tilde{\sigma}\,, \qquad \langle \sigma \rangle \equiv \frac{v}{g} \,,
\label{sigma}
\end{eqnarray}
one finds the vacuum value of $v$ by minimisation of the non-equilibrium vacuum potential
\begin{eqnarray*}
U_{\rm vac}(v)=-2v_0^2v^2+v^4 \,, \quad \frac{\partial U_{\rm vac}}{\partial v}=4 v (v^2-v_0^2)=0\,, \quad 
U_{\rm vac}(v_0)=-v_0^4\equiv \epsilon_{\rm top}(T=0)\,,
\end{eqnarray*}
in consistency with Eq.~(\ref{eps-QCD-vac}).

One can extend this model by accounting for the lightest scalar and pseudoscalar mesons $\pi^\pm$, $\pi^0$, $K^\pm$, $K^0$, 
$\bar{K}^0$, $\eta$, $\eta'$ together with $\sigma$-meson. In the simplest version of the model describing the non-interacting 
``hadron gas'' we will neglect interactions between different $\pi$, $K$, $\eta$ and $\eta'$ mesons compared to their interactions 
with the $\sigma$ meson (and hence with the QCD condensate). Such a meson gas reflects the basic properties of low-energy QCD 
at zero temperature and can be described by the effective L$\sigma$M-type chiral Lagrangian
\begin{eqnarray}
&& \mathcal{L}_{\rm eff}=\frac{1}{2}\partial_{\mu}\sigma\partial^{\mu}\sigma + 2g^2 v_0^2 \sigma^2 - g^4 \sigma^4 + \nonumber \\
&&\frac{1}{2}(\partial_{\mu}\pi_{\alpha}\partial^{\mu}\pi_{\alpha} + \partial_{\mu}\eta\partial^{\mu}\eta +
\partial_{\mu}\eta'\partial^{\mu}\eta') + \partial_{\mu}\bar{K} \partial^{\mu}K - \label{L-gas} \\
&&\frac{1}{2}\Big[2\kappa g^2 (m_u+m_d)\sigma^2 
\pi_{\alpha} \pi_{\alpha} +\frac{2}{3}\kappa g^2 (m_u+m_d+4m_s)\sigma^2 \eta^2 + \nonumber \\
&&\frac{4}{3}\kappa g^2 (m_u+m_d+m_s+\Lambda_{\rm an})\sigma^2 \eta'^2
\Big]-\kappa g^2 (m_u+m_d+2m_s)\sigma^2 \bar{K}K \,, \nonumber
\end{eqnarray}
where the relations (\ref{eps-QCD-vac}) and (\ref{masses-v0}) have been applied and the interactions are represented 
in terms of quark masses $m_{u,d,s}$, $\kappa$ defined in Eq.~(\ref{kappa}), and the gluon anomaly scale, $\Lambda_{\rm an}$. 
In analogy to the toy-model (\ref{Lsig}), the global chiral symmetry is spontaneously broken by the $\sigma$-VEV upon 
the shift in the $\sigma$-field defined in Eq.~(\ref{sigma}).

The basic idea is to generalise the ``hadron gas'' model (\ref{L-gas}) to non-zero temperatures. In what follows, we elaborate on this 
generalisation and show that it leads to phenomenologically consistent phenomena such as the condensate melting, the QCD phase transition 
and decreasing with temperature meson masses.

Let us now consider the quasi-classical system of equations for the QCD condensate and the quantum fluctuations.
Following to Carter's method \cite{Carter:1998ti, Bowman:2010zz, Mocsy:2004ab} we use an approximation in which VEVs of different fields 
are independent and equal zero for odd-point correlation functions, while even-point correlation functions can be reduced to the functions 
of smaller order. As it is discussed in Ref.~\cite{Bowman:2010zz} such an approximation corresponds to resummation of daisy and superdaisy 
diagrams. In particular, this approximation reads
\begin{eqnarray}
&& \langle\tilde{\sigma}^3\rangle=
\langle\pi_{\alpha}\pi_{\beta}\tilde{\sigma}\rangle=
\langle\eta^2\tilde{\sigma}\rangle=
\langle\eta'^2\tilde{\sigma}\rangle=
\langle\bar{K}K\tilde{\sigma}\rangle=\langle\pi_{\alpha}\tilde{\sigma}\rangle=
\langle K\tilde{\sigma}\rangle= \nonumber \\
&& \langle \bar{K}\tilde{\sigma}\rangle=
\langle \eta\tilde{\sigma}\rangle=
\langle \eta'\tilde{\sigma}\rangle=0 \,, \quad \langle\tilde{\sigma}^4\rangle=3\langle\tilde{\sigma}^2\rangle^2\,, \quad
\langle\pi_{\alpha}\pi_{\alpha}\tilde{\sigma}^2\rangle=
\langle\pi_{\alpha}\pi_{\alpha}\rangle\langle\tilde{\sigma}^2\rangle \,, \nonumber \\
&& \langle\eta^2\tilde{\sigma}^2\rangle=\langle\eta^2\rangle\langle\tilde{\sigma}^2\rangle \,, \quad
\langle\eta'^2\tilde{\sigma}^2\rangle=\langle\eta'^2\rangle\langle\tilde{\sigma}^2\rangle \,,\quad
\langle\bar{K}K\tilde{\sigma}^2\rangle=
\langle\bar{K}K\rangle\langle\tilde{\sigma}^2\rangle \,.
\label{self-cons field appr}
\end{eqnarray}
Then, using Eqs. (\ref{sigma}) and (\ref{self-cons field appr}) and the equations of motion 
one obtains the equation of state for the condensate as follows
\begin{eqnarray}
v^2&=&v_0^2-3g^2\langle\tilde{\sigma}^2\rangle - 
\frac12\kappa(m_u+m_d)\langle\pi_{\alpha} \pi_{\alpha}\rangle \nonumber \\
&-& \frac{1}{6}\kappa (m_u+m_d+4m_s)\langle\eta^2\rangle - 
\frac{1}{3}\kappa (m_u+m_d+m_s+\Lambda_{\rm an})\langle\eta'^2\rangle \nonumber \\
&-&\frac12\kappa (m_u+m_d+2m_s)\langle\bar{K}K\rangle \,. \label{eq-v}
\end{eqnarray}

Then, factorisating the operator products of the fields, one obtains the equations of motion for the scalar and pseudoscalar 
fluctuations about the ground state state corresponding to physical mesons with definite masses
\begin{eqnarray}
&&  \partial_{\mu}\partial^{\mu}\tilde{\sigma} + m_{\sigma}^2\tilde{\sigma}=0 \,, \qquad m_{\sigma}^2=8g^2v^2\,, \nonumber \\
&&  \partial_{\mu}\partial^{\mu}\pi_{\alpha} + m_{\pi}^2 \pi_{\alpha}=0\,, \qquad m_{\pi}^2=2\kappa (m_u+m_d)\mathcal{M}^2\,, \nonumber \\
&&  \partial_{\mu}\partial^{\mu}\eta + m_{\eta}^2 \eta = 0\,, \qquad m_{\eta}^2=\frac{2}{3}\kappa (m_u+m_d+4m_s)\mathcal{M}^2\,, \label{e-o-m} \\
&&  \partial_{\mu}\partial^{\mu}\eta' + m_{\eta'}^2 \eta'=0\,, \qquad m_{\eta'}^2=\frac{4}{3}\kappa (m_u+m_d+m_s+\Lambda_{\rm an})\mathcal{M}^2\,, \nonumber \\ 
&&  \partial_{\mu}\partial^{\mu}K + m_{K}^2 K=0\,, \qquad m_{K}^2=\kappa (m_u+m_d+2m_s)\mathcal{M}^2\,, \nonumber
\end{eqnarray}
where $v^2$ is found in Eq.~(\ref{eq-v}), and
\begin{eqnarray}
\mathcal{M}^2\equiv v^2+g^2\langle\tilde{\sigma}^2\rangle \,. 
\label{eq-M}
\end{eqnarray}
The $\sigma$ quartic coupling $g$ can be found from the vacuum value of the $\sigma$-meson mass (at $T=0$) 
known phenomenologically as the scalar $f_0(500)$ state \cite{Agashe:2014kda}
\begin{eqnarray}
m_{\sigma\rm{(vac)}}^2=8g^2v_0^2\,, \qquad m_{\sigma\rm{(vac)}}\sim 400-550\,\rm{MeV} \,,
\end{eqnarray}
and the condensate amplitude $v_0$ known from Eq.~(\ref{v0}). Comparing the physical masses of $\pi$, $K$, $\eta$ and $\eta'$ 
mesons in the low-symmetric phase found in Eq.~(\ref{e-o-m}) to those in the vacuum given by Eq.~(\ref{masses}), we notice that the difference 
is accounted via the change in the order parameter as $v_0 \to \mathcal{M}$.

%%%%%%%%%%%%%%%%%%%%%%%%%
\section{Thermodynamics of the meson plasma}
\label{Sec:Thermodynamics}
%%%%%%%%%%%%%%%%%%%%%%%%%

%%%%%%%%%%%%%%%%%%%%
\subsection{Generating functional}
\label{Sec:functional}
%%%%%%%%%%%%%%%%%%%%

In the case of zero chemical potential, the free energy of the meson plasma is given by the expectation value 
of the trace of the spatial part of the energy-momentum tensor $\mathcal{T}_{i}^{j}$, $i,j=1,2,3$ (for the full energy-momentum 
tensor of this model, see Eq.~(\ref{T})). Using Eqs.~(\ref{sigma}), (\ref{self-cons field appr}), (\ref{e-o-m}) 
as well as the results for the expectation values of field derivatives as was discussed in Appendix~\ref{App:temp int}, 
it is straightforward to construct the generating functional which is the free energy density written as a function of 
the order parameter $v$, temperature $T$ and meson masses $m_{\sigma}$, $\mathcal{M}$ as follows
\begin{eqnarray}\nonumber
&& \mathcal{F}(T,v,m_{\sigma},\mathcal{M}) \equiv \frac{1}{3}\langle\mathcal{T}_{i}^{i}\rangle = 
-\frac{1}{3}\big[J_2(T,m_{\sigma})+3J_2(T,m_{\pi}) +  \\
&& J_2(T,m_{\eta}) + 4J_2(T,m_K) + J_2(T,m_{\eta'})\big] + U(v,m_{\sigma},\mathcal{M})\,, \label{F3}
\end{eqnarray}
where $J_2(T,m)$ is defined in Eq.~(\ref{J2}), the physical meson masses (except for $m_{\sigma}$), 
$m_{\phi}$, $\phi\equiv \{\pi,\,K,\,\eta,\,\eta'\}$ are related to $\mathcal{M}$ as given by Eq.~(\ref{e-o-m}), and
\begin{eqnarray}
&& U(v,m_{\sigma},\mathcal{M}) = \frac{m^4_{\sigma}}{64\pi^2}\ln\frac{m^2_{\sigma}}{\sqrt{e}\,m^2_{\sigma\rm{(vac)}}} +
\frac{3m^4_{\pi}}{64\pi^2}\ln\frac{m^2_{\pi}}{\sqrt{e}\,m^2_{\pi\rm{(vac)}}} + \nonumber \\
&& \frac{m^4_{\eta}}{64\pi^2}\ln\frac{m^2_{\eta}}{\sqrt{e}\,m^2_{\eta\rm{(vac)}}} + 
\frac{4m^4_{K}}{64\pi^2}\ln\frac{m^2_{K}}{\sqrt{e}\,m^2_{K\rm{(vac)}}} +
\frac{m^4_{\eta'}}{64\pi^2}\ln\frac{m^2_{\eta'}}{\sqrt{e}\,m^2_{\eta'\rm{(vac)}}} - \label{U} \\
&& \frac{m_{\sigma}^2}{2g^2}(\mathcal{M}^2 - v^2) - 2v_0^2\mathcal{M}^2 + v^4 +
6v^2(\mathcal{M}^2 - v^2) + 3(\mathcal{M}^2 - v^2)^2 \,, \nonumber
\end{eqnarray}
in terms of the vacuum values of the order parameter $v_0$ and the meson masses $m_{\phi\rm{(vac)}}$, 
and the base of natural logarithm $e$. Indeed, $T,\,v,\, m_{\sigma},\,\mathcal{M}$ are the independent
variables of the functional of state, the generating functional $\mathcal{F}$, containing the full thermodynamic 
information about the system, its vacuum and the spectrum of perturbations.

The extremum conditions of the generating functional $\mathcal{F}$ over $v,\, m_{\sigma},\,\mathcal{M}$ read
\begin{eqnarray}
&& \frac{\partial \mathcal{F}}{\partial v}\Big|_{T,m_{\sigma},\mathcal{M}} \equiv 
\frac{v}{g^2}(m^2_{\sigma}-8g^2v^2)=0\,, \label{extr v} \\
&& \frac{\partial \mathcal{F}}{\partial m_{\sigma}}\Big|_{T,v,\mathcal{M}} \equiv 
- \frac{m_{\sigma}}{g^2}\Big\{\mathcal{M}^2-v^2-g^2\Big[J_{1}(T,m_{\sigma}) +
\frac{m_{\sigma}^2}{16\pi^2}\ln\frac{m_{\sigma}^2}{m_{\sigma\rm{(vac)}}^2}\Big]\Big\}=0\,, \label{extr ms} \\
&& \frac{\partial \mathcal{F}}{\partial \mathcal{M}}\Big|_{T,v,m_{\sigma}} \equiv 4\mathcal{M}
\Big\{v^2 - v_0^2 - \frac{1}{4g^2}(m_{\sigma}^2 - 8g^2v^2) +
3(\mathcal{M}^2 - v^2) + \Delta\Big\}=0 \,, \label{extr M}
\end{eqnarray}
where
\begin{eqnarray}
&& \Delta(T,\mathcal{M}) =\frac{3}{2}\kappa (m_{u}+m_{d})
\Big[J_{1}(T,m_{\pi})+\frac{1}{16\pi^2}m_{\pi}^2\ln\frac{m_{\pi}^2}{m_{\pi\rm{(vac)}}^2}\Big]+ \nonumber \\
&&\frac{1}{6}\kappa (m_{u}+m_{d}+4m_{s})
\Big[J_{1}(T,m_{\eta})+\frac{1}{16\pi^2}m_{\eta}^2\ln\frac{m_{\eta}^2}{m_{\eta\rm{(vac)}}^2}\Big]+ \label{Delta}\\
&&\kappa (m_{u}+m_{d}+2m_{s})
\Big[J_{1}(T,m_{K})+\frac{1}{16\pi^2}m_{K}^2\ln\frac{m_{K}^2}{m_{K\rm{(vac)}}^2}\Big]+ \nonumber \\
&&\frac{1}{3}\kappa (m_{u}+m_{d}+m_{s}+\Lambda_{\rm an})
\Big[J_{1}(T,m_{\eta'})+\frac{1}{16\pi^2}m_{\eta'}^2\ln\frac{m_{\eta'}^2}{m_{\eta'\rm{(vac)}}^2}\Big]\,. \nonumber 
\end{eqnarray}
The first relation (\ref{extr v}) coincides with the $\sigma$-mass relation in Eq.~(\ref{e-o-m}). The other two 
relations (\ref{extr ms}) and (\ref{extr M}) can be resolved w.r.t. $v^2$ and $\mathcal{M}$ leading to the condensate equation of state
\begin{eqnarray}
v^2=v_0^2-3g^2\Big[J_{1}(T,m_{\sigma})+\frac{1}{16\pi^2}m_{\sigma}^2\ln\frac{m_{\sigma}^2}{m_{\sigma\rm{(vac)}}^2}\Big] - \Delta 
\label{eq v1}
\end{eqnarray}
and hence to the physical $\sigma$-mass at finite temperature while the other meson masses are expressed via
\begin{eqnarray}
\mathcal{M}^2=v^2+g^2\Big[J_{1}(T,m_{\sigma})+\frac{1}{16\pi^2}m_{\sigma}^2\ln\frac{m_{\sigma}^2}{m_{\sigma\rm{(vac)}}^2}\Big]\,.
\label{eq M2}
\end{eqnarray}
Note, Eqs.~(\ref{eq v1}) and (\ref{eq M2}) can be directly obtained from their definitions (\ref{eq-v}) and (\ref{eq-M}) by an explicit 
calculation of expectation values of the fields, as it was also shown in \cite{Bowman:2010zz,Mocsy:2004ab}.

Alternatively, using Eqs.~(\ref{extr ms}) and (\ref{extr M}) one could resolve all the meson masses in terms of temperature $T$ and 
the order parameter $v$ as $m_{\sigma}=m_{\sigma}(T,v)$, $m_{M}\propto \mathcal{M}=\mathcal{M}(T,v)$. So, after a simple substitution 
of the latter solutions the generating functional is transformed to the so-called non-equilibrium Landau functional
\begin{eqnarray}
\mathcal{F}_{\rm NE}(T,v)\equiv \mathcal{F}(T,v,m_{\sigma}(T,v),\mathcal{M}(T,v))\,.
\label{FTv}
\end{eqnarray}
Finally, resolving the $\sigma$-mass relation
\begin{eqnarray} \label{m-sig}
m_{\sigma}(T,v)^2=8g^2v^2 \,,
\end{eqnarray}
w.r.t. $v=v(T)$ and substituting this result into Eq.~(\ref{FTv}), one arrives at the equilibrium Landau functional
\begin{eqnarray}
\mathcal{F}_{\rm E}(T)\equiv \mathcal{F}_{\rm NE}(T,v(T))\,,
\label{FT}
\end{eqnarray}
corresponding to the ordinary free energy.

Applying the above extrema solutions $m_{\sigma}=m_{\sigma}(T,v), \, \mathcal{M}=\mathcal{M}(T,v) $, 
the stability condition of the low-symmetry phase (in the case of a non-trivial VEV $v\not=0$) reads
\begin{eqnarray}
\frac{d^2\mathcal{F}_{\rm NE}}{dv^2} = -16v^2\frac{\alpha - \frac{1}{4}\delta}{\alpha} \geq 0 \,, \quad 
\alpha \equiv 1+g^4\delta\Big[J_{0}(T,m_{\sigma}) + J_{0\rm{(vac)}}(m_{\sigma})\Big] \,,
\label{stab}
\end{eqnarray}
where $J_{0\rm{(vac)}}(m)$ is given in Eq.~(\ref{Jvac}), and
\begin{eqnarray}
&& \delta(T,\mathcal{M})=6 - 3\kappa^2(m_u+m_d)^2
\Big(J_{0}(T,m_{\pi})+J_{0\rm{(vac)}}(m_{\pi})\Big) -\nonumber \\
&& \frac{1}{9}\kappa^2(m_u+m_d+4m_s)^2
\Big(J_{0}(T,m_{\eta}) + J_{0\rm{(vac)}}(m_{\eta})\Big) - \label{delta} \\
&& \frac{4}{9}\kappa^2(m_u+m_d+m_s+\Lambda_{\rm an})^2
\Big(J_{0}(T,m_{\eta'}) + J_{0\rm{(vac)}}(m_{\eta'})\Big) - \nonumber \\
&& \kappa^2(m_u+m_d+2m_s)^2\Big(J_{0}(T,m_{K}) + 
J_{0\rm{(vac)}}(m_{K})\Big) \,. \nonumber
\end{eqnarray}
In Eq.~(\ref{stab}), the equality 
\begin{eqnarray}
\label{Tc}
\frac{d^2\mathcal{F}_{\rm NE}}{dv^2}\Big|_{T=T_c} = 0
\end{eqnarray}
defines the critical (phase transition) temperature $T=T_c$.

A numerical analysis of Eq.~(\ref{stab}) for acceptable values of the quartic coupling $g$ shows that for small temperatures $T\ll T_c$
both the numerator and the denominator in Eq.~(\ref{stab}) are positively-definite and finite. With the increase of $T\to T_c$ 
the denominator increases and remains finite while the numerator decreases and vanishes at $T=T_c$. This observation enables us
to write down the stability condition for the low-symmetry phase of the meson plasma as
\begin{eqnarray}
\alpha - \frac{1}{4}\delta \leq 0\,, \quad m_{\sigma}=m_{\sigma}(T,v)\,,
\label{stab-1}
\end{eqnarray}
%%%%%%%%%%%%%%%%%%%%%%%%%%%%%%%%%%%%%%%%
\begin{figure*}[!h]
\begin{minipage}{0.45\textwidth}
 \centerline{\includegraphics[width=1\textwidth]{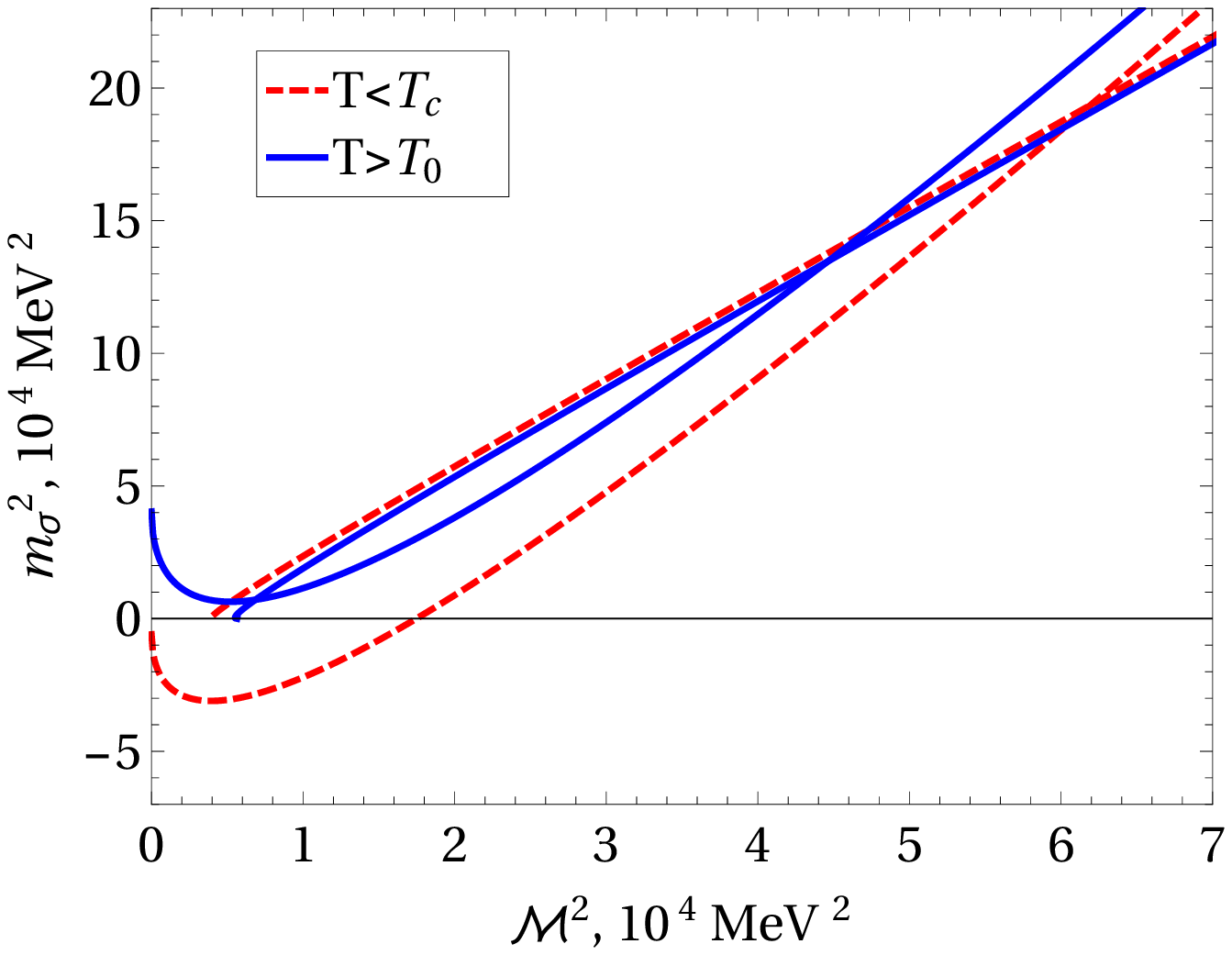}}
\end{minipage}
\begin{minipage}{0.455\textwidth}
 \centerline{\includegraphics[width=1\textwidth]{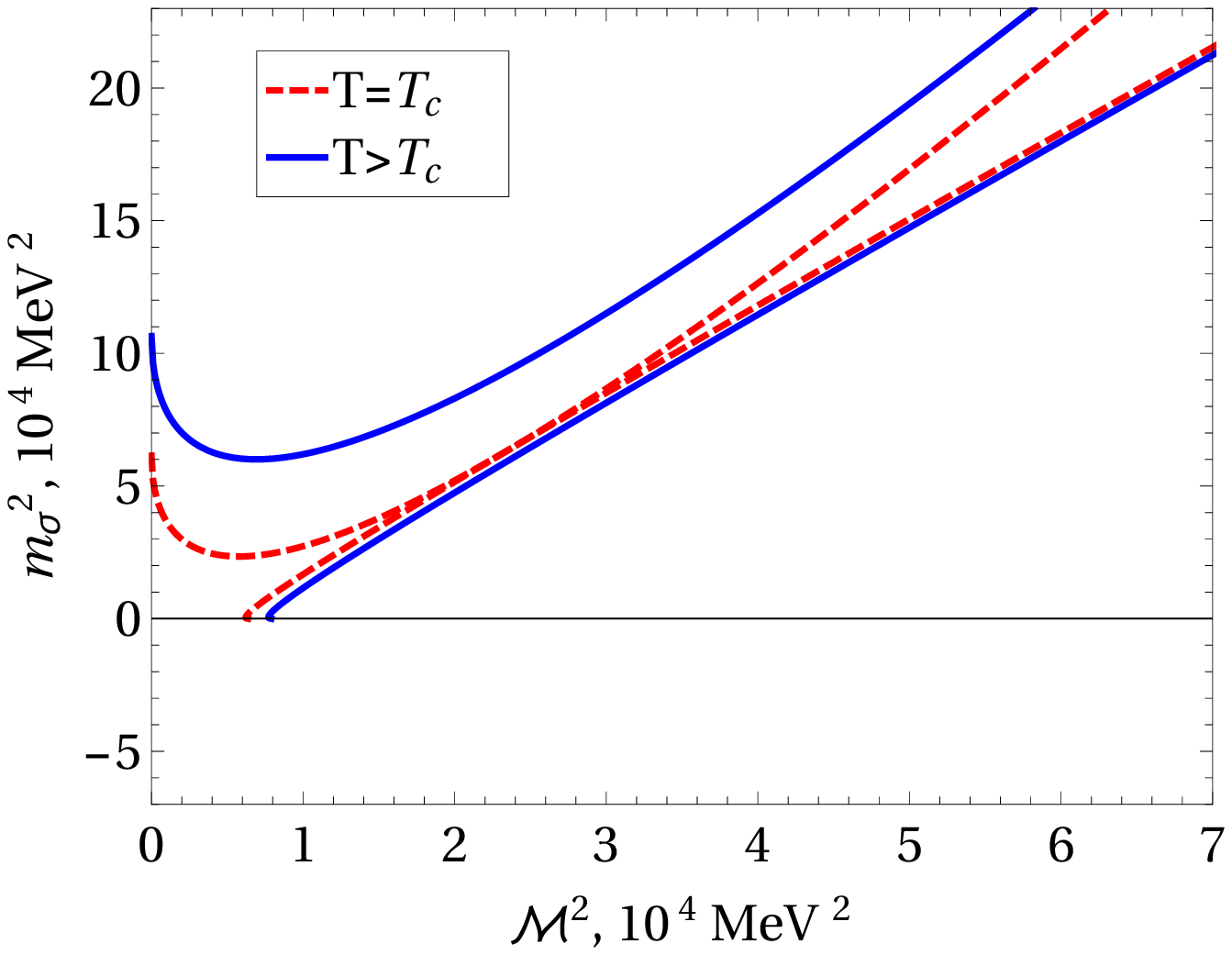}}
\end{minipage}
\caption{The graphical analysis of the system (\ref{msM}).}
\label{fig:msM1}
\end{figure*}
%%%%%%%%%%%%%%%%%%%%%%%%%%%%%%%%%%%%%%%%

Provided that $\Delta$ given in Eq.~(\ref{Delta}) does not explicitly depend on the order parameter $v$, i.e. $\Delta=\Delta(T,\mathcal{M})$,
it is convenient to exclude $v^2$ from the system of equations (\ref{extr v}), (\ref{extr ms}) and (\ref{extr M}) reducing them to the following 
two equations
\begin{eqnarray} \nonumber
&& m_{\sigma}(T,\mathcal{M})^2=-4g^2v_0^2+12g^2\mathcal{M}^2+4g^2\Delta(T,\mathcal{M})\,, \\
&& \mathcal{M}(T,m_{\sigma})^2=\frac{m_{\sigma}^2}{8g^2}+g^2\Big[J_{1}(T,m_{\sigma})+\frac{1}{16\pi^2}
m_{\sigma}^2\ln\frac{m_{\sigma}^2}{m_{\sigma\rm{(vac)}}^2}\Big]\,.
\label{msM}
\end{eqnarray}
that can be now solved e.g. graphically for different temperatures as shown in Fig.~\ref{fig:msM1}. At low temperatures $T<T_c$ (red lines) shown 
in Fig.~\ref{fig:msM1} (left panel) there is only one solution where the lines intersect for which the stability condition (\ref{stab-1}) is satisfied. When 
temperature grows and reaches $T=T_0$ the second solution appears for lower mass values than for the first solution $m_{\sigma(2)}<m_{\sigma(1)}$, 
$\mathcal{M}_{(2)}<\mathcal{M}_{(1)}$. Despite the positivity of the meson mass spectrum, the stability condition (\ref{stab-1}) does not satisfy 
for the second solution so it is found to be unstable. With further growth of temperature the meson masses corresponding to the first stable solution 
decrease and at $T=T_c$ the stable solution joins the unstable one as presented in Fig.~\ref{fig:msM1} (red dashed curves on the right panel) i.e. the curves 
touch in a single point where the corresponding derivatives match
\begin{eqnarray}
\frac{dm_{\sigma}^2(\mathcal{M}^2)}{d\mathcal{M}^2}\Big|_{T=T_c}=\Big(\frac{d\mathcal{M}^2(m_{\sigma})}{dm_{\sigma}^2}\Big)^{-1}\Big|_{T=T_c} \,.
\end{eqnarray}
which is a direct analog of the condition on the critical temperature $T_c$ (\ref{Tc}) where the phase transition occurs. At higher temperatures $T>T_c$ 
there are no solutions of the system (\ref{msM}). Provided that at the critical point the values $v_c$, $m_{\sigma(c)}$, $\mathcal{M}_{c}$ are non-zero
the phase transition is of the first order.
%%%%%%%%%%%%%%%%%%%%%%%%%%%%%%%%%%%%%%%
\begin{figure*}[!h]
 \centerline{\includegraphics[width=0.5\textwidth]{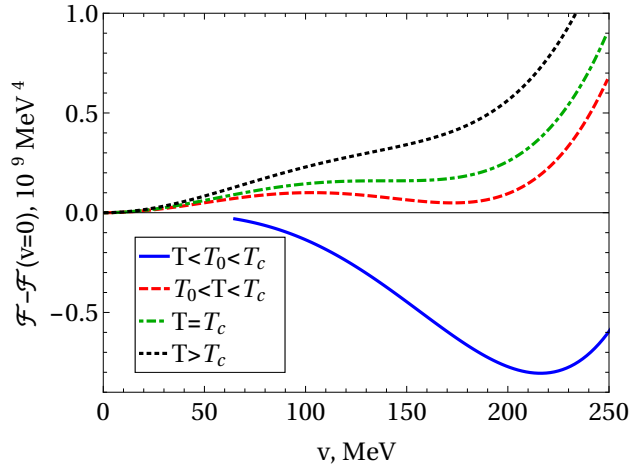}}
\caption{The non-equilibrium Landau functional, $\mathcal{F}_{\rm NE}(T,v)$, as a function of $v$ 
for different temperatures (for low $v$ and $T_0<T<T_c$, blue line, there is no finite solution of Eqs.~(\ref{extr ms}) and (\ref{extr M}) 
which defines the functional (\ref{FTv})).}
\label{fig:F}
\end{figure*}
%%%%%%%%%%%%%%%%%%%%%%%%%%%%%%%%%%%%%%%

Another way of representing the corresponding dynamics is by considering the Landau functional (\ref{FTv}) directly looking for the equilibrium states 
numerically in different characteristic temperature regions. In Fig.~\ref{fig:F} one notices that at low temperature $T<T_c$ there is a stable phase with 
$v(T)\neq 0$ where the extremum conditions (\ref{extr v})--(\ref{extr M}) are satisfied (blue solid line). As the temperature $T$ grows, the order 
parameter $v=v(T)$ decreases. At some specific value $T=T_0$ a new maximum appears (red dashed line). For $T=T_c$, the minimum for $v\neq 0$ 
joins the maximum and both non-trivial extrema disappear such that the low-symmetric phase becomes unstable (green dot-dashed line). In the case of 
$T>T_c$, there is only one global minimum at $v=0$ where the stability conditions are satisfied (black dotted line). As the value 
of the order parameter $v$ changes abruptly and stepwise from $v=v_c$ to $v=0$, the first-order phase transition indeed occurs in the considered model at $T=T_c$ 
as was already indicated above. The generic analysis of the minimisation conditions (\ref{extr v}), (\ref{extr ms}), (\ref{extr M}) and (\ref{stab}) enables us to reconstruct 
the behaviour of the basic thermodynamic quantities as functions of temperature. It will be shown below that, indeed, this behaviour corresponds to the first-order phase 
transition at $T=T_c$. 

The characteristic observables of the meson plasma can be built based upon the equilibrium Landau potential (free energy) (\ref{FT}), $\mathcal{F}_{\rm E}(T)$.
Namely, pressure $p(T)$, entropy density $\sigma(T)$, energy density $\epsilon$, and heat capacity $c_V$ are found as
\begin{eqnarray}
p(T)=-\mathcal{F}_{\rm E}(T) \,, \quad \sigma(T)=-\frac{d}{dT}\mathcal{F}_{\rm E}(T)\,, \quad \epsilon=\mathcal{F}_{\rm E}+T\sigma \,, \quad
c_V=T\frac{d\sigma}{dT}=\frac{d\epsilon}{dT}\,,
\label{quantaties}
\end{eqnarray}
respectively. These quantities as given in explicit analytic form in Appendix~\ref{App:obs}. The speed of sound squared is then given by
\begin{eqnarray}
u^2=\Big(\frac{dp}{d\epsilon}\Big)_{\sigma} = \frac{dp/dT}{d\epsilon/dT}=\frac{\sigma}{c_V}\,.   
\label{speed}
\end{eqnarray}

%%%%%%%%%%%%%%%%%%%%%%%%%%%%%%
\subsection{Zero condensate phase}
\label{Sec:zero-v}
%%%%%%%%%%%%%%%%%%%%%%%%%%%%%%

At temperatures above $T_c$, the system is expected to reach a deconfined state where the condensate is melted.
According to Ref.~\cite{Nahrgang:2013xaa,Turko:2014jta} mesons can survive near (pseudo-)critical point as metastable states, 
and, hence, it is reasonable to ask about the properties of mesons at $T>T_c$. 

Of course, above the critical temperature $T>T_c$, the characteristic degrees of freedom in the considering system change since hadrons 
get deconfined into quarks and gluons, and the condensate melts down, thus, approaching the zero-condensate phase. Such a deconfined 
phase has to be considered separately based upon a new generating functional written in terms of relevant degrees of freedom in the deconfined 
plasma different from that in Eq.~(\ref{F3}). But, as an initial step one can consider mesonic fluctuations in such an exotic phase without quarks, gluons 
and with nearly-melted condensate.

The existence of this phase follows from Eq.~(\ref{extr v}). According to Eq.~(\ref{extr v}) the non-equilibrium Landau functional (\ref{FTv}) can 
have an extremum at $v=0$ already at temperatures below $T_c$. In this case the system of equations (\ref{extr M}) and (\ref{extr ms}) turns into
\begin{eqnarray}
&& \mathcal{M}^2 - g^2\Big[J_{1}(T,m_{\sigma}) + \frac{1}{16\pi^2}m_{\sigma}^2
\ln\frac{m_{\sigma}^2}{m_{\sigma\rm{(vac)}}^2}\Big]=0 \,,   \nonumber\\ 
&& -v_0^2 - \frac{1}{4g^2}m_{\sigma}^2+3\mathcal{M}^2+\Delta = 0 \label{zero eq}
\end{eqnarray}
that has to have a real solution $\{m_{\sigma},\,\mathcal{M}\}$ for the considering temperature.
The stability condition for such a phase (\ref{stab}) corresponds to a positively-definite $\sigma$-meson mass
\begin{eqnarray}
m_{\sigma}^2>0\,,
\label{stab zero}
\end{eqnarray}
which is satisfied automatically as long as the system (\ref{zero eq}) has a solution. As was mentioned above in the context of Fig.~\ref{fig:F}, the zero condensate phase 
becomes metastable at certain $T_0<T_c$ where the $\sigma$ meson mass vanishes $m_{\sigma}(T_0)=0$, while the pseudoscalar mass parameter
\begin{eqnarray}
\mathcal{M}(T_0)^2 = \frac{g^2}{12}T_0^2
\end{eqnarray}
is finite. For vanishing $v(T_0)=0$, the system of equations (\ref{extr M}) and (\ref{extr ms}) transforms to the transcendental algebraic equation
\begin{eqnarray}
aT_0^2 + bT_0^2\, {\rm ln}\frac{g^2T_0^2}{12v_0^2} - v_0^2=0\,, \label{eq-T0}
\end{eqnarray}
whose solution reads
\begin{eqnarray}
T_0=\frac{v_0}{\sqrt{b W[\frac{e^{a/b}g^2}{12b}]}}\,, \label{T0}
\end{eqnarray}
where $W$ is the Lambert function, and the parameters $a,\,b$ are found as
\begin{eqnarray}
a&=&\frac{g^2}{4} + \kappa\Bigg\{\frac{3}{2}(m_u + m_d)\widetilde{J}_{1}\bigg(\frac{\kappa(m_u + m_d)g^2}{6}\bigg) \nonumber \\ 
&+& \frac{1}{6}(m_u+m_d+4m_s)\widetilde{J}_{1}\bigg(\frac{\kappa(m_u+m_d+4m_s)g^2}{18}\bigg) \nonumber \\ 
&+&(m_u+m_d+2m_s)\widetilde{J}_{1}\bigg(\frac{\kappa(m_u+m_d+2m_s)g^2}{12}\bigg) \nonumber \\ 
&+&\frac{1}{3}(m_u+m_d+m_s+\Lambda_{\rm an})\widetilde{J}_{1} 
\bigg(\frac{\kappa(m_u+m_d+m_s+\Lambda_{\rm an})g^2}{9}\bigg)\Bigg\} \,, \nonumber \\
b&=&\frac{g^2\kappa^2}{192\pi^2}\bigg\{3(m_u + m_d)^2 + \frac{1}{9}(m_u + m_d + 4m_s)^2 \nonumber \\ 
&+& (m_u + m_d + 2m_s)^2 + \frac{4}{9}(m_u+m_d+m_s+\Lambda_{\rm an})^2\bigg\} \,. \nonumber
\end{eqnarray}
Here, the function $\widetilde{J}_1$ is defined by Eq.~(\ref{J wave}). For a wide range of parameters, $b\ll a$, such that to a good
approximation
\begin{eqnarray}
T_0\approx \frac{v_0}{\sqrt{a}}\,. \label{T0-approx}
\end{eqnarray}

At some temperature $T_1$ the zero-condensate phase stabilises which corresponds to the situation, when two minima, $v=0$ and $v\neq0$ of non-equilibrium 
functional $F_{\rm NE}(T,v)$ become equal, and beyond $T_c$ it is the only stable phase (see Fig.~\ref{fig:F}). In the minimum $v\neq0$, the non-equilibrium 
functional equals to the equlibrium one $F_E$. Then, temperature $T_1$ can be defined from the following equation
\begin{eqnarray}
F_{\rm NE}(T_1,0)=F_{\rm E}(T_1)\,. \label{T1}
\end{eqnarray}

For e.g.~$m_{\sigma({\rm vac})}=500\,{\rm MeV}$ one has $T_1=430\,{\rm MeV}$. The thermodynamic observables in the zero-condensate phase can be 
obtained by using relations (\ref{quantaties}) and (\ref{speed}), while the corresponding equilibrium generating functional is $F_{\rm E0}(T)=F_{\rm NE}(T,0)$ 
(c.f.~Appendix~\ref{App:obs}). In particular, pressure, entropy density and energy density are described by the same formulae (\ref{pressure}), (\ref{entropy}), 
(\ref{energy}), but computed for $v=0$ and with $m_{\sigma}$, $\mathcal{M}$ following from the corresponding solution of Eq.~(\ref{zero eq}).

The chiral phase transition to the zero-condensate phase occurs for some temperature between $T_1$, when the zero-condensate phase stabilizes, and $T_c$, 
when the phase with the finite condensate becomes unstable. If such a transition happens at $T=T_1$, then the latent heat is given by
\begin{eqnarray}
\Delta \varepsilon =\varepsilon_{v=0}-\varepsilon_{v\neq 0}=1.0\times 10^{10}\,{\rm MeV}^4 \,. \label{T1}
\end{eqnarray}

%%%%%%%%%%%%%%%%%%%%%%%%%%%%%%%
\subsection{Thermal fluctuations in the meson plasma}
\label{Sec:th-fluc}
%%%%%%%%%%%%%%%%%%%%%%%%%%%%%%%

Let us briefly discuss thermal fluctuations in the system ``QCD condensate + meson plasma'' in the vicinity of the phase transition for low-symmetric phase. 
The probability of a fluctuation of the order parameter $\Delta v$ behave as follows
\begin{eqnarray}
w \sim {\rm exp}\Big[ -\frac{\Delta F}{T}\Big]\sim {\rm exp}\Big[-\frac{(\Delta v)^2}{2\langle(\Delta v)^2\rangle}\Big]\,,\label{wf}
\end{eqnarray}
where $F=\mathcal{F}_{\rm NE}V$ is a full free energy (non-equilibrium Landau functional) in the meson plasma occupying volume $V$. The relative 
mean square fluctuation of the order parameter can be obtained by Taylor-expanding $\Delta F$ in Eq.~(\ref{wf}) up to the second order such that
\begin{eqnarray}
\Big\langle\frac{(\Delta v)^2}{v^2}\Big\rangle = -\frac{T}{16v^4 V}\,\frac{\alpha}{\alpha-\frac{1}{4}\delta} \,, \label{dvv}
\end{eqnarray}
where $\alpha$ and $\delta$ are defined in Eqs.~(\ref{stab}) and (\ref{delta}), respectively. The mean square amplitude of the meson 
mass fluctuations can be obtained by expressing the $\Delta m_{\sigma}$ and $\Delta \mathcal{M}$ fluctuations through $\Delta v$ 
resulting in
\begin{eqnarray}
\Big\langle\frac{(\Delta m_{\sigma})^2}{m_{\sigma}^2}\Big\rangle =
-\frac{T \delta^2 g^4}{4m_{\sigma}^4 V}\, \frac{1}{\alpha(\alpha-\frac14\delta)} \,, \qquad
\Big\langle\frac{(\Delta \mathcal{M})^2}{\mathcal{M}^2}\Big\rangle =
-\frac{T}{16\mathcal{M}^4 V}\, \frac{1}{\alpha(\alpha-\frac14\delta)} \,. \label{dm}
\end{eqnarray}
When the system warms up with $T\to T_c$, the term $\alpha-\delta/4\to 0$ vanishes according to Eq.~(\ref{Tc}). This means that fluctuations of 
the order parameter and masses singularly increase as $\propto (\alpha-\delta/4)^{-1/2}$ in the system ``searching'' for a new stable state.

%%%%%%%%%%%%%%%%
\section{Numerical results}
\label{Sec:num}
%%%%%%%%%%%%%%%%

%%%%%%%%%%%%%%%%
\subsection{Mass spectrum}
%%%%%%%%%%%%%%%%

Let us investigate the evolution of meson masses with temperature below and above the critical temperature $T_c$ numerically. The $\sigma$-meson 
mass at zero temperature is phenomenologically uncertain and belongs to a wide region $400\, {\rm MeV} \lesssim m_{\sigma{\rm (vac)}} \lesssim 
550\, {\rm MeV}$ that corresponds to a variation of the quartic coupling squared within the interval $0.26<g^2<0.48$. The latter variation in the coupling 
has then been employed for finding the typical values of the critical temperature $T_c$, the meson mass parameters at this temperature such as $m_{\sigma(c)}$ 
and $\mathcal{M}_{c}$, as well as the strength of the first-order phase transition given terms of the ratios $v_c/T_c$ or $m_{\sigma(c)}/T_c$. The results are 
shown in Table~\ref{Tab:strength}.
%%%%%%%%%%%%%%%%
\begin{table}[ht]
\centering
\caption{The first-order phase transition strength as a function of the quartic coupling squared $0.26<g^2<0.48$.}
\begin{tabular}{|c|c|c|c|c|c|c|c|c|}
\hline
$m_{\sigma{\rm (vac)}}$, MeV & $g^2$ & $m_{\sigma(c)}$, MeV & $\mathcal{M}_{c}$, MeV & $v_c$, MeV & $T_c$, MeV  & $v_c/T_c$ & 
$m_{\sigma(c)}/T_c$ & $\mathcal{M}_{c}/T_c$\\
\hline
400 & 0.26 & 202 & 150 & 141 & 443 & 0.319 & 0.46 & 0.34\\
425 & 0.29 & 214 & 151 & 141 & 442 & 0.319 & 0.48 & 0.34\\
450 & 0.32 & 226 & 151 & 140 & 441 & 0.319 & 0.51 & 0.34\\
475 & 0.36 & 238 & 152 & 140 & 439 & 0.319 & 0.54 & 0.35\\
500 & 0.40 & 250 & 153 & 140 & 438 & 0.320 & 0.57 & 0.35\\
525 & 0.44 & 263 & 153 & 140 & 437 & 0.321 & 0.60 & 0.35\\
550 & 0.48 & 276 & 154 & 141 & 435 & 0.323 & 0.63 & 0.35\\
\hline
\end{tabular}
\label{Tab:strength}
\end{table}
%%%%%%%%%%%%%%%%

One notices that the strength of the first-order phase transition increases for larger vacuum $\sigma$-meson masses.

\begin{figure*}[!h]
 \centerline{\includegraphics[width=0.6\textwidth]{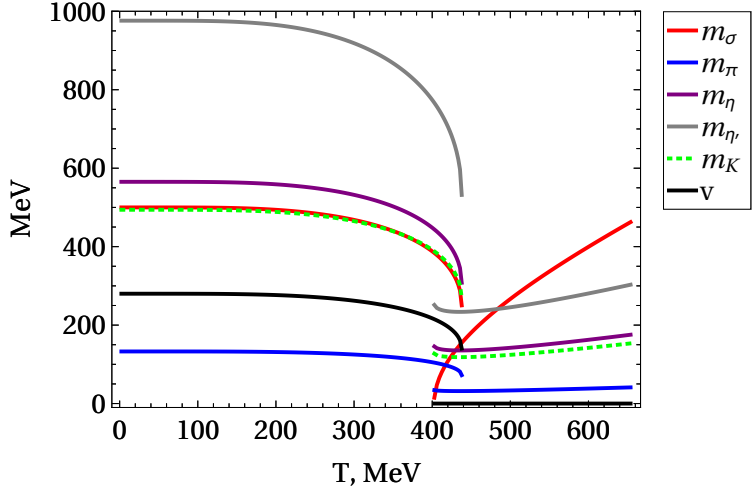}}\vspace{1cm}
\caption{The meson masses and the amplitude of the condensate as functions of temperature for $g^2=0.4$.}
\label{fig:masses}
\end{figure*}

The thermal evolution of the meson masses and the condensate amplitude is shown in Fig.~\ref{fig:masses} 
for $m_{\sigma{\rm (vac)}}=500\,{\rm MeV}$ and $g^2=0.4$ corresponding to $T_c = 438\,{\rm MeV}$ 
(see Table~\ref{Tab:strength}). The masses of all the mesons and the condensate amplitude in the plasma 
decrease with $T$ at $T<T_c$. Such a characteristic behavior is different from that in other approaches \cite{Mocsy:2004ab, Blaschke:2014zsa, 
Bowman:2010zz} where the $\sigma$-meson mass also decreases but the pion mass remains either finite or increases. The variation of 
$m_{\sigma{\rm (vac)}}$ (and $g^2$) shown in Table~\ref{Tab:strength} only slightly affects the critical temperature as well as the thermal 
evolution of the meson masses. The fact that $T_c$ turns out to be larger than the conventional estimates $150-200 \, {\rm MeV}$ 
\cite{Bazavov:2014pvz, Bazavov:2011nk} can be related with the too narrow meson spectrum accounted for in the considering model 
(for example, one could check that in the homogeneous $\sigma$-meson plasma the critical temperature is much higher). 
Also, such a value of the critical temperature can be further reduced by inclusion of fermions (quarks and baryons) as was noticed 
earlier in Ref.~\cite{Mocsy:2004ab}.

Above $T_c$ the masses of pseudoscalar mesons unevenly fall down to small values and then start to slowly increase while the $\sigma$-meson mass 
falls down to zero and then increase relatively fast. Such behaviour is also a result of the ``hadron gas'' approximation and it differs from the conventional 
L$\sigma$M or PNJL results \cite{Mocsy:2004ab, Blaschke:2014zsa, Bowman:2010zz}. This observation can be interesting since one could notice from Fig.~\ref{fig:masses} 
that $m_{\sigma}$ is larger than $2m_{\pi}$ almost everywhere. This means that $\sigma$ will decay to a pair of pions, i.e.~$\sigma\to\pi^+\pi^-$ and 
$\sigma\to\pi^0\pi^0$, which was kinematically forbidden in other approaches \cite{Mocsy:2004ab, Blaschke:2014zsa, Bowman:2010zz} where 
$m_{\sigma}\to m_{\pi}$ at large $T>T_c$.
%%%%%%%%%%%%%%%%
%%%%%%%%%%%%%%%%

As was discussed in Sec.~\ref{Sec:zero-v}, the zero condensate phase $v=0$ becomes metastable at a certain $T_0<T_c$ that corresponds to a vanishing 
$\sigma$-meson mass $m_{\sigma}(T_0)=0$. The values of $T_0$, $\mathcal{M}(T_0)$ as well as $m_{\sigma(c)}$, $\mathcal{M}_{c}$ in the 
zero condensate phase and temperature of phase stabilization $T_1$ are collected in Table~\ref{Tab:zero-phase}.
%%%%%%%%%%%%%%%%%%%%%%%%%%%
\begin{table}[ht]
\centering
\caption{Characteristics of the zero condensate phase.}
\begin{tabular}{|c|c|c|c|c|c|c|c|}
\hline
$m_{\sigma{\rm (vac)}}$, MeV & $g^2$ & $T_0$, MeV & $T_1$, MeV & $(T_c-T_0)/T_c$ & $\mathcal{M}(T_0)$, MeV & 
$m_{\sigma}(T_c)$, MeV  & $\mathcal{M}(T_c)$, MeV \\
\hline
400 & 0.26 & 401 & 435 & 0.10 & 58.5 & 139 & 55.5 \\
425 & 0.29 & 402 & 433 & 0.09 & 62.2 & 143 & 58.5 \\
450 & 0.32 & 402 & 432 & 0.09 & 66.0 & 147 & 61.4 \\
475 & 0.36 & 402 & 431 & 0.08 & 69.7 & 151 & 64.3 \\
500 & 0.40 & 402 & 430 & 0.08 & 73.3 & 155 & 67.0 \\
525 & 0.44 & 402 & 429 & 0.08 & 76.9 & 158 & 69.8 \\
550 & 0.48 & 401 & 428 & 0.08 & 80.5 & 162 & 72.5 \\
\hline
\end{tabular}
\label{Tab:zero-phase}
\end{table}
%%%%%%%%%%%%%%%%%%%%%%%%%%%

We find that the width of the phase coexistence region is about $10\%$ of the critical temperature and slightly decreases with $g^2$ while $T_0$ is non-monotonic. 

%%%%%%%%%%%%%%%%%%%%%%%%%%%%%%%%%%%%%%%%%%%
\subsection{Pressure, energy density and equation of state of the meson plasma}
%%%%%%%%%%%%%%%%%%%%%%%%%%%%%%%%%%%%%%%%%%%

\begin{figure*}[!h]
\begin{minipage}{0.45\textwidth}
 \centerline{\includegraphics[width=1\textwidth]{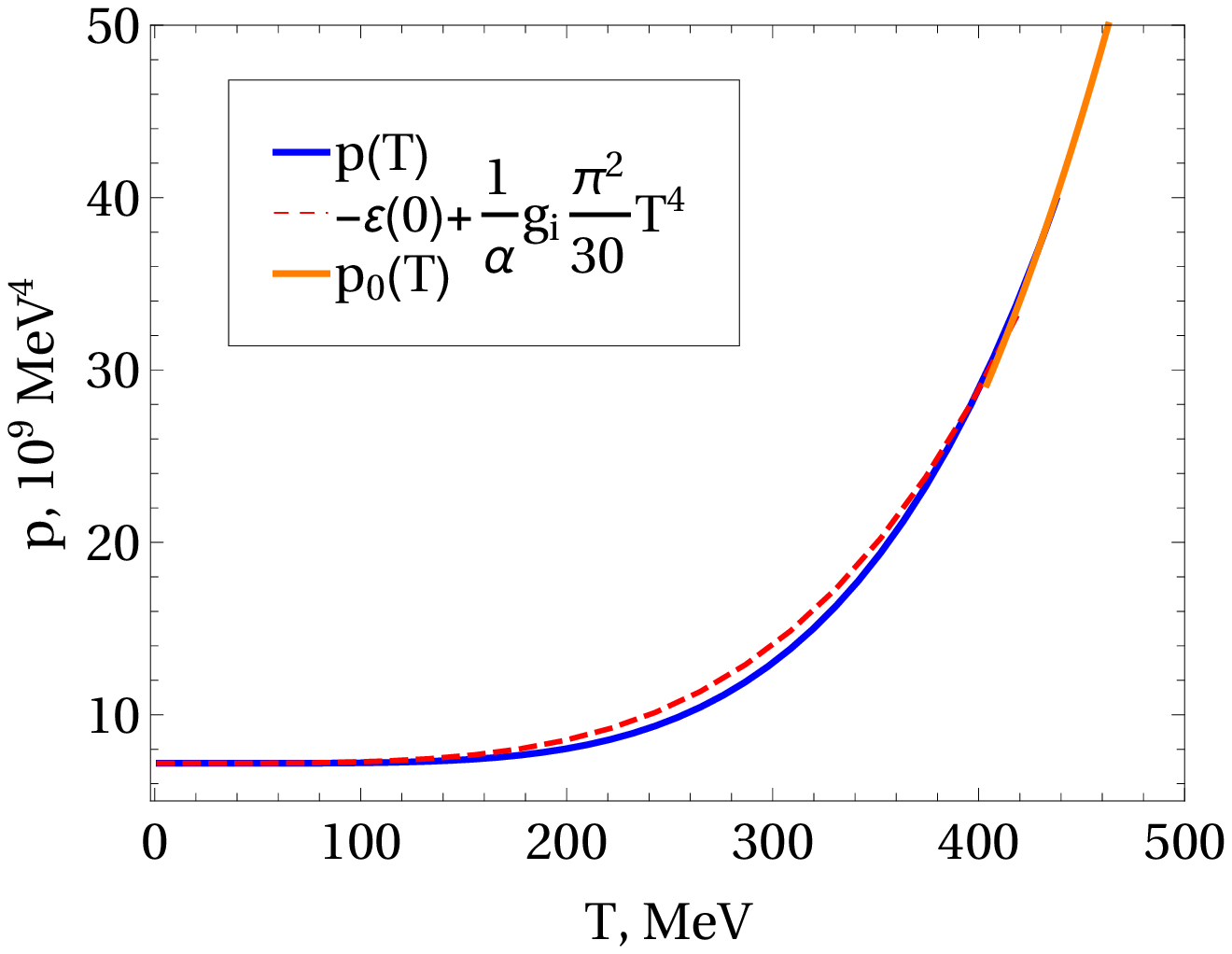}}
\end{minipage}
\begin{minipage}{0.45\textwidth}
 \centerline{\includegraphics[width=1\textwidth]{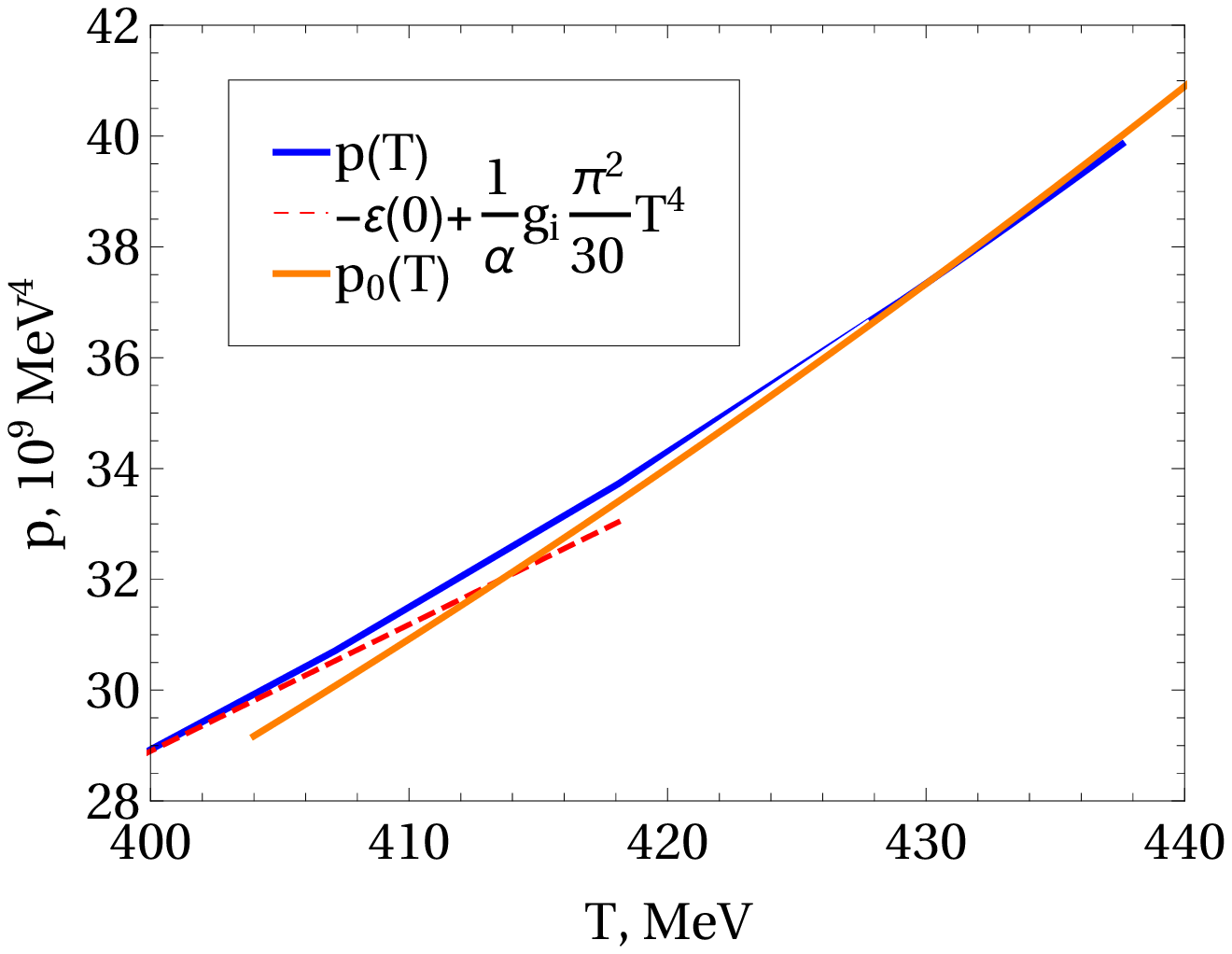}}
\end{minipage}
\caption{Thermal evolution of pressure, $p(T)$ corresponds to $v\neq 0$ phase and $p_0(T)$ to zero-condensate phase.}
\label{fig:pressure-and-energy}
\end{figure*}

The formulas for energy density pressure are provided by Eqs.~(\ref{energy}) and (\ref{pressure}), respectively, and their thermal 
evolution is shown in Fig.~\ref{fig:pressure-and-energy}. The results weakly depend on $g$-coupling values.
%%%%%%%%%%%%%%%%%%%%%%%%%%%%%%%%%
%%%%%%%%%%%%%%%%%%%%%%%%%%%%%%%%%

Taking the low-temperature limits $T\to 0$ of Eqs.~(\ref{pressure}) and (\ref{energy}) using the relations for the temperature integrals in Eqs.~(\ref{J wave}) 
and (\ref{zero J}), we find that only vacuum contributions in pressure and density remain. The net QCD vacuum energy density $\epsilon^{\rm QCD}_{\rm vac}$ turns out 
to be negative 
\begin{eqnarray}
\label{QCD-vac-eps}
\epsilon^{\rm QCD}_{\rm vac} &\equiv& \epsilon(T=0) = \epsilon_{\rm top}+\epsilon^{\rm had}_{\rm vac} = - v_0^4 - 
\frac{1}{128\pi^2}\big(m_{\sigma{\rm (vac)}}^4 + 3m_{\pi{\rm (vac)}}^4 \nonumber \\
&+& m_{\eta{\rm (vac)}}^4+4m_{K{\rm (vac)}}^4 + m_{\eta'{\rm (vac)}}^4\big)\simeq -7\times 10^9\,{\rm MeV}^4\,.
\end{eqnarray}
Apparently, it gets two contributions: topological (quark-gluon) condensate density, $\epsilon_{\rm top} = - v_0^4$, that has been introduced 
earlier in Eq.~(\ref{eps-QCD-vac}), and an addition term that corresponds to the perturbative hadronic vacuum $\epsilon^{\rm had}_{\rm vac}$
due to the regularised contributions from meson fluctuations. The net QCD vacuum density and pressure at $T=0$ satisfy the vacuum equation
of state, namely,
\begin{eqnarray}
\epsilon(T=0)=-p(T=0)
\end{eqnarray}
as they should. The hadronic vacuum contribution $\epsilon^{\rm had}_{\rm vac}$ turns out to be rather small 
$\epsilon^{\rm had}_{\rm vac}/\epsilon^{\rm QCD}_{\rm vac}\approx 0.15$ and enters with negative sign. 

\begin{figure*}[!h]
\begin{minipage}{0.45\textwidth}
 \centerline{\includegraphics[width=1\textwidth]{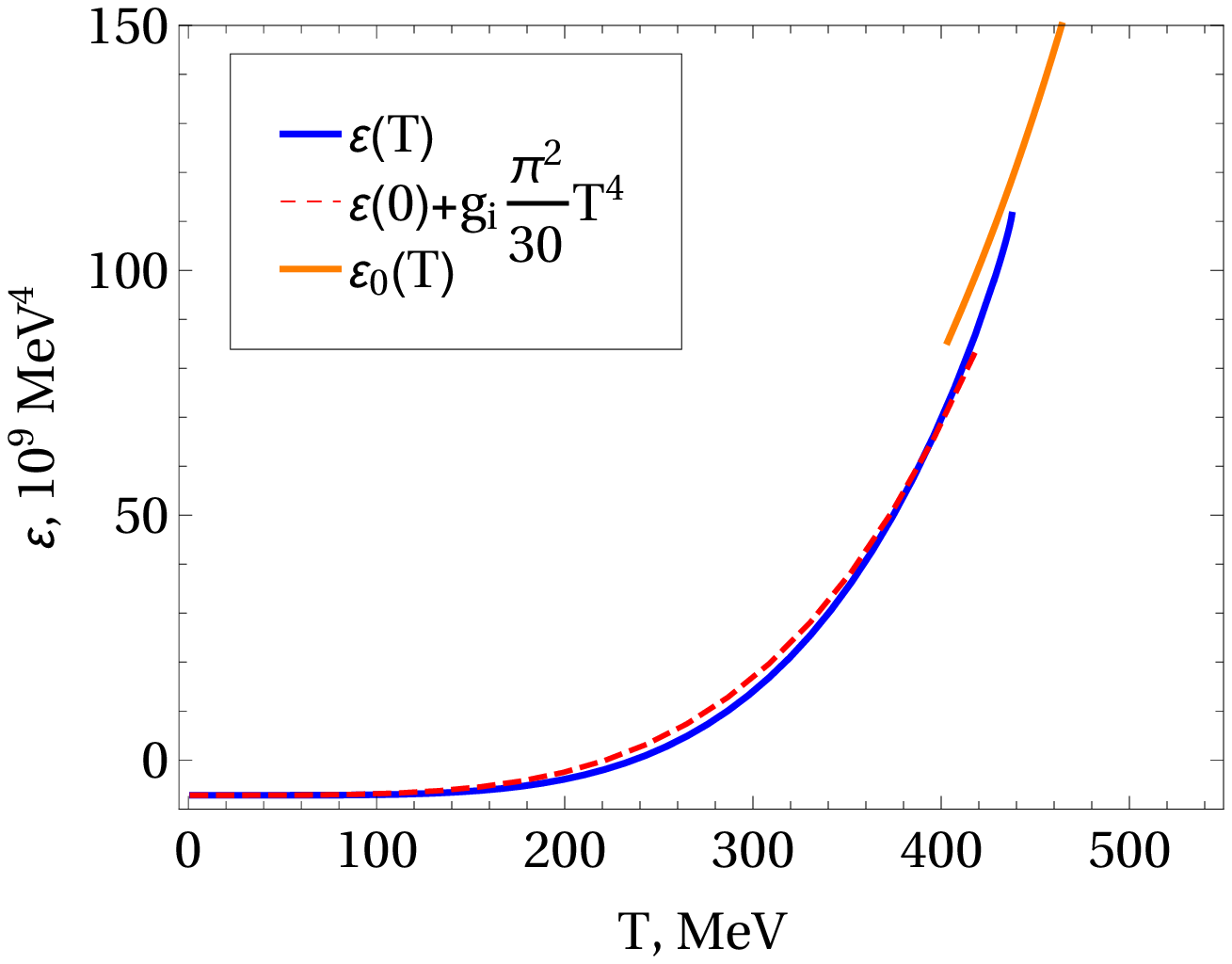}}
\end{minipage}
\begin{minipage}{0.45\textwidth}
 \centerline{\includegraphics[width=1\textwidth]{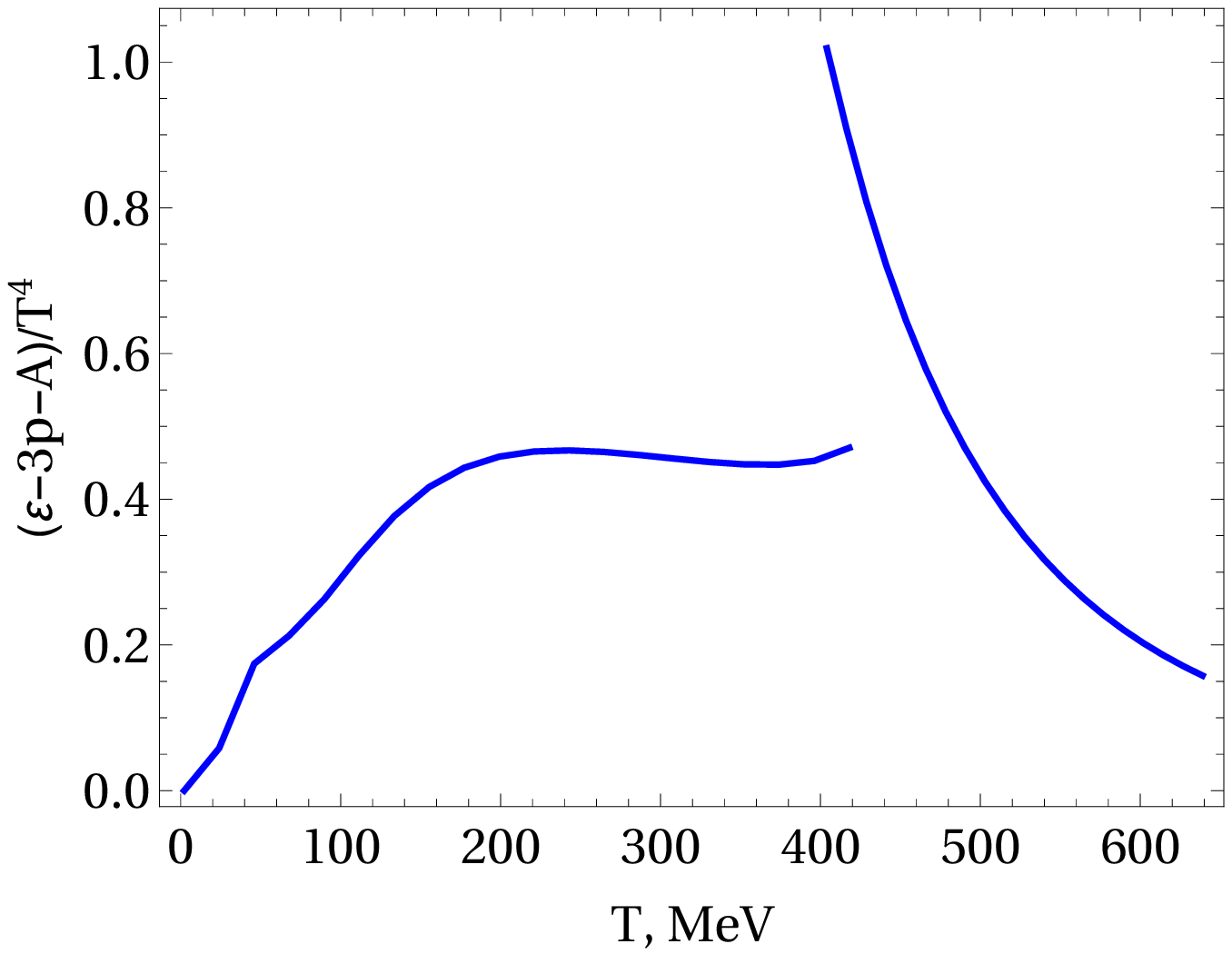}}
\end{minipage}
\caption{Thermal evolution of the energy density (left), where $\varepsilon(T)$ corresponds to $v\neq 0$ phase and 
$\varepsilon_0(T)$ -- to the zero-condensate phase, and the equation of state (right).}
\label{fig:state-equation}
\end{figure*}

With the increase of temperature, both pressure and energy density grow due to positive particle contributions. The total energy density 
vanishes $\epsilon(T_{\epsilon=0})=0$ at $T_{\epsilon=0}\simeq 237\,{\rm MeV}$ for $m_{\sigma(vac)}=500\,{\rm MeV}$. Interestingly enough, as can 
be seen in Figs.~\ref{fig:pressure-and-energy} and \ref{fig:state-equation}, the resulting form of pressure and energy density profiles can be approximated 
by the corresponding relations built upon the constant negative vacuum contribution and the massless bosonic plasma term $\propto T^4$. A good fit can 
be obtained for the coefficient $\alpha\simeq 3.5$ and for the effective number of relativistic degrees of freedom $g_i\simeq 9$. At $T>T_c$, the energy density 
increases unevenly while pressure changes continuously. At $T=T_1$ pressure is the same in both phases.

The equation of state $(\epsilon(T)-3p(T)-A)/T^4$, where $A=\varepsilon(T=0)-3p(T=0)$ is a vacuum contribution, changes from the vacuum value which 
corresponds to zero up to positive values as shown in Fig.~\ref{fig:state-equation}. It then decreases in the zero-condensate phase at $T>T_c$.
%%%%%%%%%%%%%%%%%%%%%%%%%%%%%%%%%%%
%%%%%%%%%%%%%%%%%%%%%%%%%%%%%%%%%%%

%%%%%%%%%%%%%%%%%%%%%%%%%%%%%%%%%%%
\subsection{Entropy density, heat capacity and speed of sound}
%%%%%%%%%%%%%%%%%%%%%%%%%%%%%%%%%%%

\begin{figure*}[!h]
\begin{minipage}{0.32\textwidth}
 \centerline{\includegraphics[width=1\textwidth]{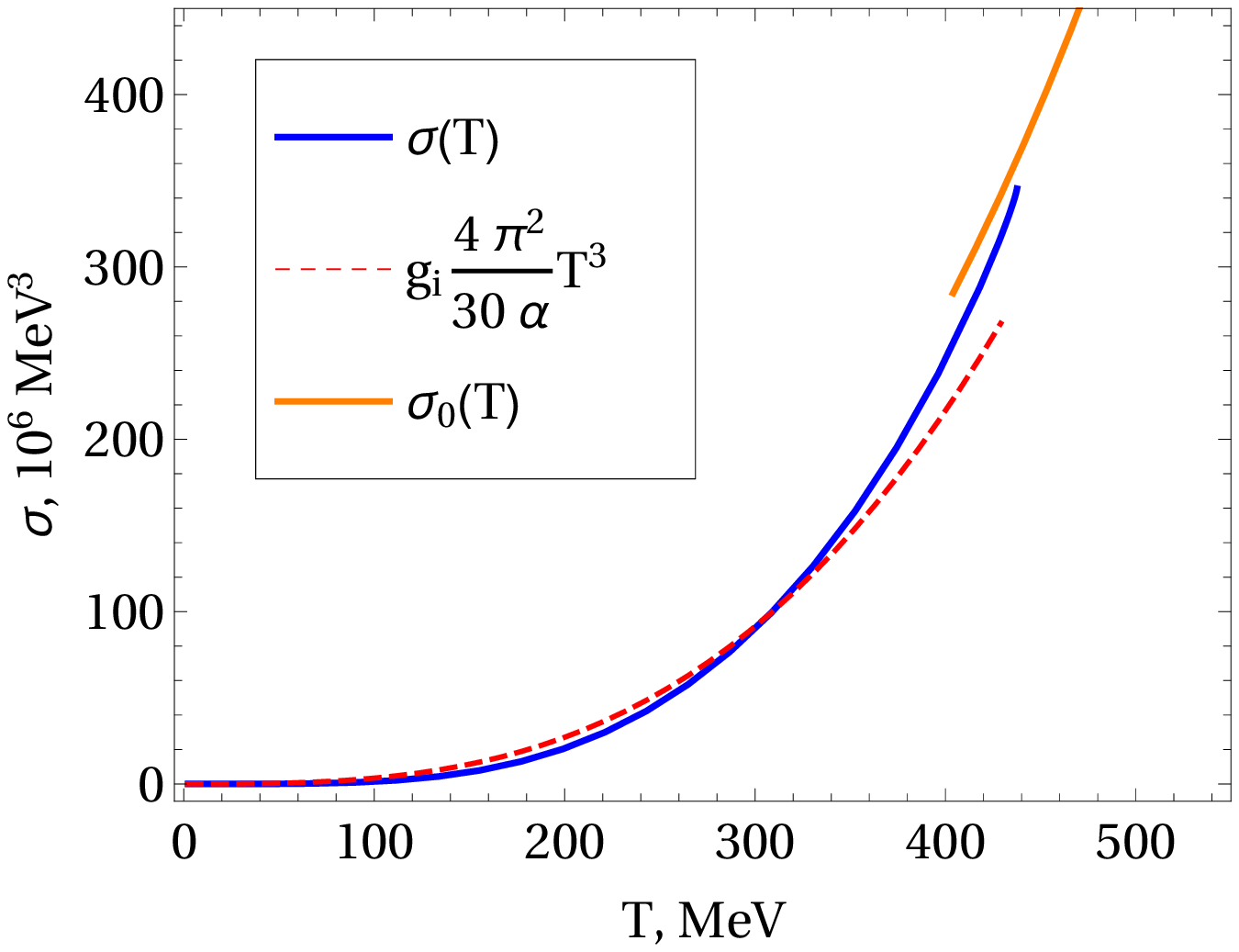}}
\end{minipage}
\begin{minipage}{0.32\textwidth}
 \centerline{\includegraphics[width=1\textwidth]{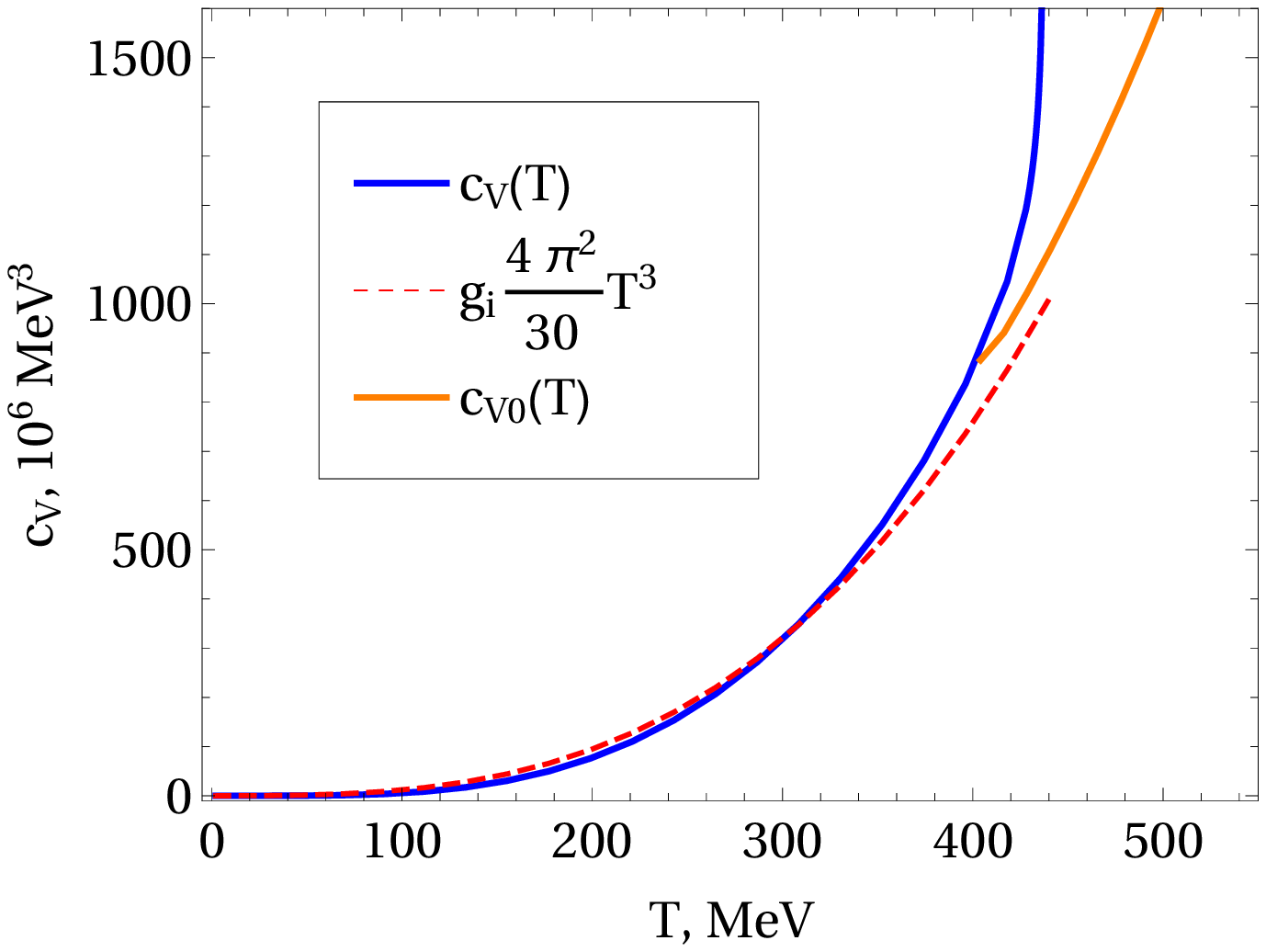}}
\end{minipage}
\begin{minipage}{0.32\textwidth}
 \centerline{\includegraphics[width=1\textwidth]{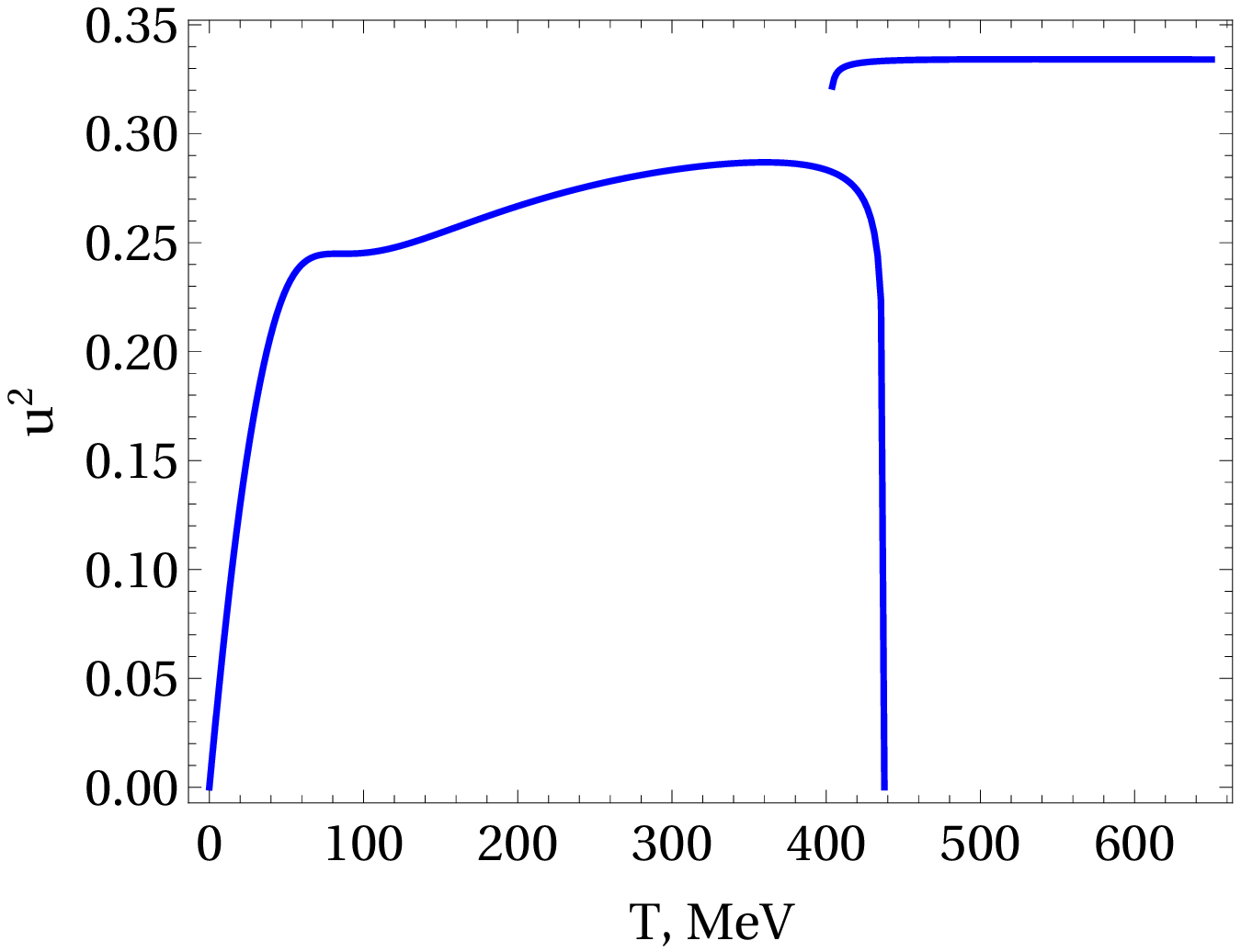}}
\end{minipage}
\caption{The entropy density (left), the heat capacity (middle) and 
the speed of sound squared (right) of the meson plasma.}
\label{fig:entropy-capacity-speed}
\end{figure*}

Now consider other thermodynamical characteristics of the meson plasma such as the entropy density, 
the heat capacity and the speed of sound squared which are given by the formulas (\ref{entropy}), 
(\ref{capacity}) and (\ref{speed}), respectively. The corresponding numerical results are shown 
in Fig.~\ref{fig:entropy-capacity-speed} for both phases.
%%%%%%%%%%%%%%%%%%%%%%%%%%
%%%%%%%%%%%%%%%%%%%%%%%%%%

At small temperatures $T\ll T_c$, the entropy density, the heat capacity and the speed of sound squared behave as follows
\begin{eqnarray}
&& \sigma(T\to 0)=\frac{3}{(2\pi)^{3/2}} m_{\pi{\rm (vac)}}^{5/2}\sqrt{T}e^{-m_{\pi{\rm (vac)}}/T}\,, \nonumber\\ 
&& c_V(T\to 0)=\frac{3}{(2\pi)^{3/2}} m_{\pi{\rm (vac)}}^{7/2}\frac{1}{\sqrt{T}}e^{-m_{\pi{\rm (vac)}}/T}\,, \label{Tto0}\\ 
&& u^2(T\to 0)=\frac{T}{m_{\pi{\rm (vac)}}}\,, \nonumber
\end{eqnarray}
respectively, and turn to zero at $T=0$ as expected. The latter relations show that the low-temperature plasma is dominated by 
the lightest states, namely, $\pi$-mesons. 

At large temperatures $T\to T_c$, the behavior of these quantities corresponds to the first-order phase transition. Indeed, the entropy 
density $\sigma$ increases but remains finite and the heat capacity $c_V$ has a singularity at $T=T_c$ as follows from Eqs.~(\ref{stab-1}), 
(\ref{capacity}) and (\ref{dmdT}), while the speed of sound squared $u^2$ turns to zero according to Eq.~(\ref{speed}). Note, the latter 
remains roughly constant in a wide range of temperatures $50\,{\rm MeV}\lesssim T\lesssim 400\,{\rm MeV}$ and beyond $T_c$.

%%%%%%%%%%%%%%%%%%%%%%%%%%%
\section{QCD vacuum in cosmology}
\label{Sec:Cosmology}
%%%%%%%%%%%%%%%%%%%%%%%%%%%

In cosmology, the presence of vacuum condensates with negative energy density leads to a contradiction as the right hand side of the Friedmann equations 
has to be positive for the flat universe. More precisely, if we introduce a negative QCD vacuum contribution then the universe will necessarily collapse 
at some point in its history. Indeed, after the phase transition at $T\sim T_c$, the net QCD energy density is positive, see Fig.~\ref{fig:state-equation} (left panel). 
Then, as the universe expands and the energy density of matter components decreases, approximately as $1/a^n$ ($n=4$ for radiation and $n=3$ for non-relativistic matter). 
The negative QCD vacuum contribution $\epsilon^{\rm QCD}_{\rm vac}$ found in Eq.~(\ref{QCD-vac-eps}) behaves as a cosmological constant and does not depend 
on the scale factor. Thus, at some moment the negative-valued condensate totally compensates a positive contribution from matter such that the total energy 
vanishes. At this moment the Hubble constant $H=\dot{a}/a$ also vanishes for the flat Friedmann universe while $\ddot{a}<0$ as the pressure is always 
positive, see Fig.~\ref{fig:pressure-and-energy} (right panel). This means that the universe begins to contract at this time and collapses very fast during 
the time period of the order of a microsecond as will be shown below. Such a situation has nothing to do with reality as the modern large-scale universe would not be formed 
in this case so the negative QCD condensate must be eliminated in some way, e.g. by invoking an additional positively-definite cosmological constant (see below).

Let us discuss more quantitatively the behavior of the flat universe filled with the meson plasma in the model considered above neglecting all other possible 
contributions apart from QCD, for simplicity. The Friedmann equations for the scale factor in physical time $a=a(t)$ are as follows
\begin{eqnarray}
\frac{3}{\varkappa} \frac{\dot{a}^2}{a^2}=\epsilon \,,\qquad 
\frac{d}{dt}(\epsilon a^4) - \frac{d a}{dt}a^3(\epsilon - 3p)=0\,, 
\label{Friedmann}
\end{eqnarray}
where $\epsilon=\epsilon(T)$ and $p=p(T)$ are the total energy density and pressure of the ``meson plasma + QCD condensate'' medium at $T<T_c$ given by 
Eqs.~(\ref{energy}) and (\ref{pressure}), respectively. The corresponding equation for the trace of the energy-momentum tensor 
can be written as
\begin{eqnarray}
\frac{6}{\varkappa}\Big(\frac{\ddot{a}}{a}+\frac{\dot{a}^2}{a^2}\Big)=\epsilon - 3p \,. 
\label{Friedmann trace}
\end{eqnarray}
and in what follows we suppose that the first order phase transition from/to the meson plasma phase happens at $T=T_c$ during heating/cooling of the universe. 
A more accurate analysis of the phase transition domain requires taking into account the QGP phase, of course, and for this reason we do not consider the non-equilibrium
processes of bubbles' creation as well as the superheated and overcooled states in this work.

\begin{figure*}[!h]
\begin{minipage}{0.45\textwidth}
 \centerline{\includegraphics[width=1\textwidth]{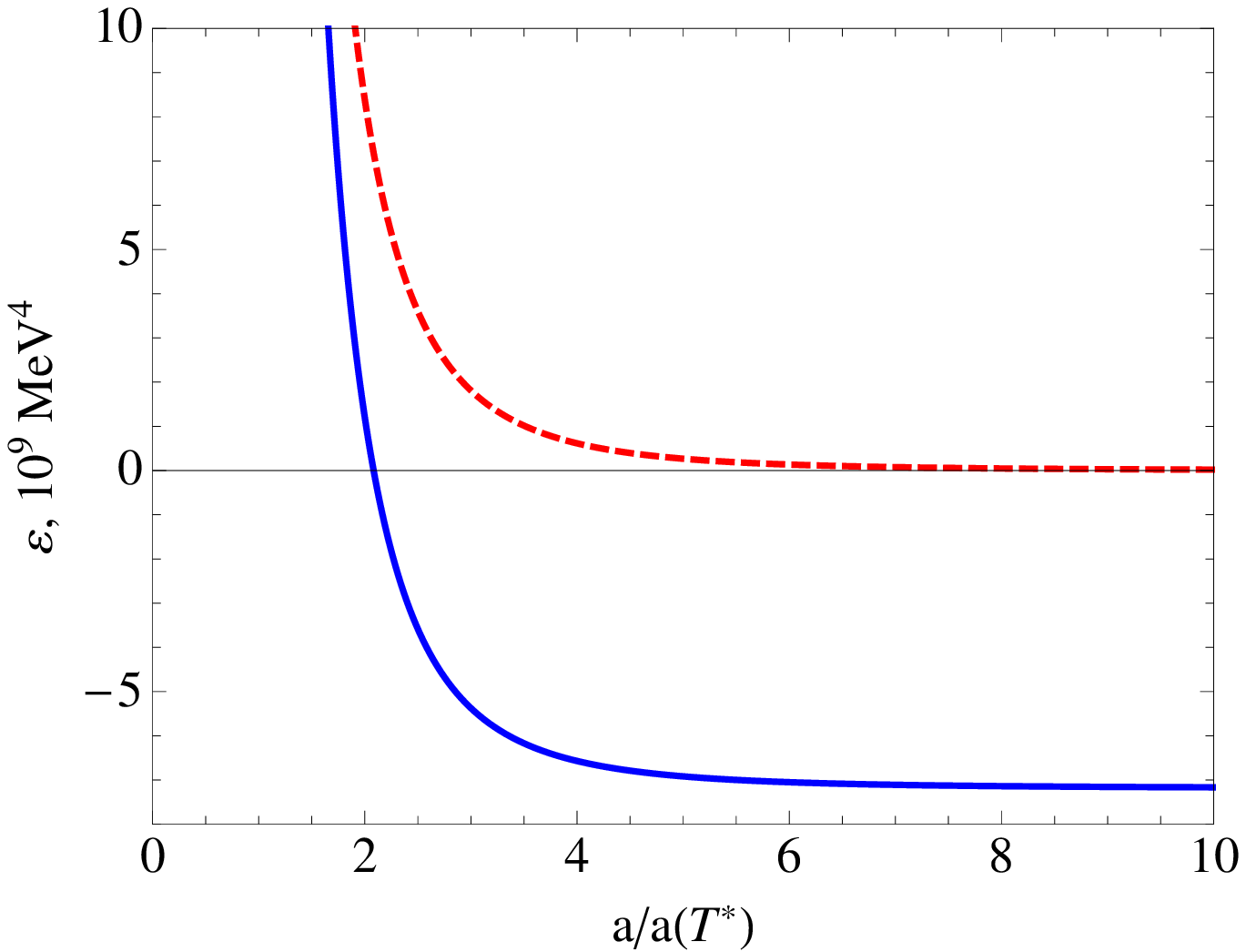}}
\end{minipage}
\begin{minipage}{0.45\textwidth}
 \centerline{\includegraphics[width=1\textwidth]{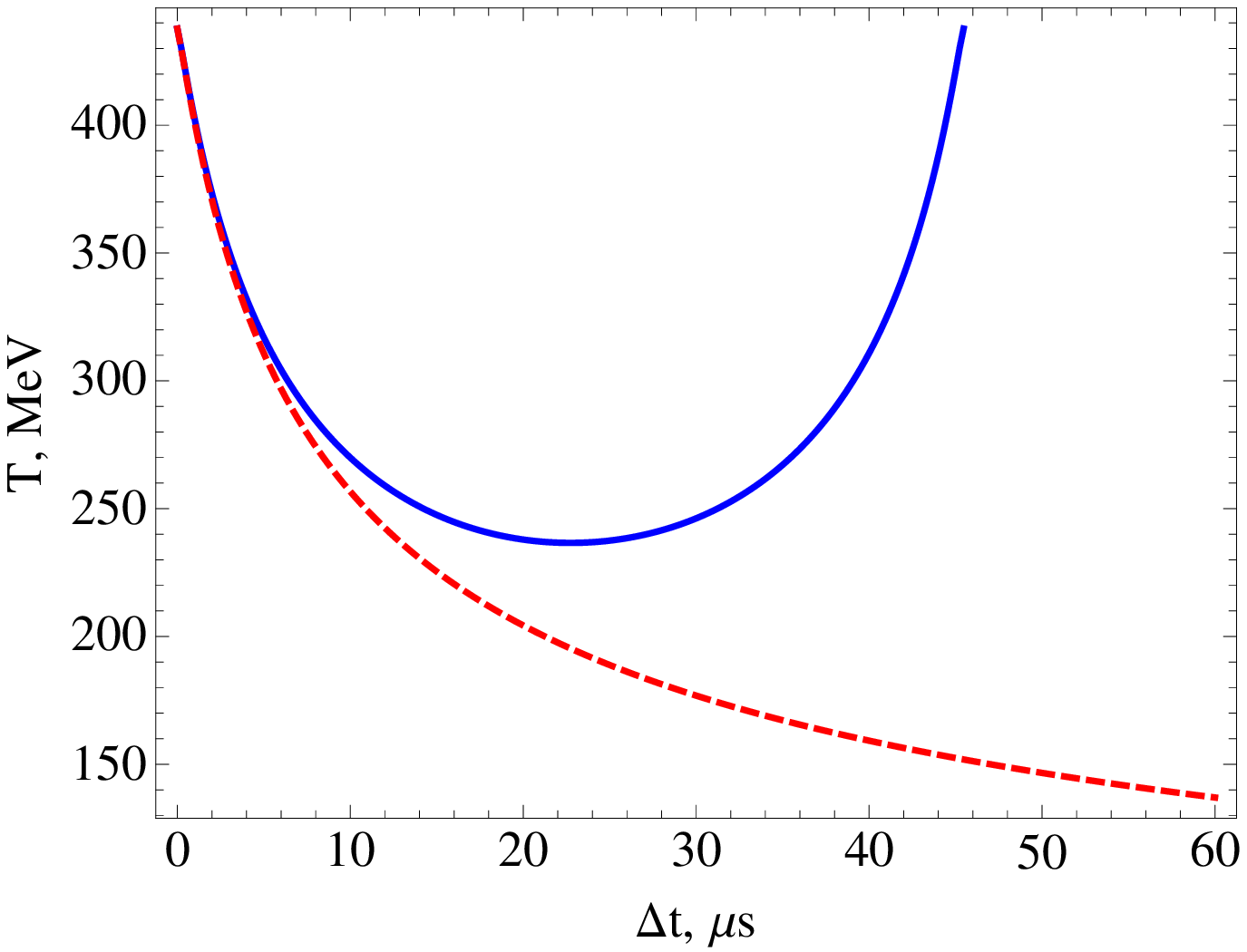}}
\end{minipage}
\caption{The energy density as a function of the scale factor in the meson plasma model (solid line) and for the same model but with additional 
positive QCD-scale $\Lambda$-term (dashed line) is shown on the left panel. The temperature as a function of time (solid line) and for the same model but with additional 
positive QCD-scale $\Lambda$-term (dashed line) is presented on the right panel.}
\label{fig:ea-Tt}
\end{figure*}

The second equation in Eq.~(\ref{Friedmann}) enables us to express the scale factor through the entropy density $\sigma=\sigma(T)$. Indeed, 
for any $T$ and $T^*$ below $T_c$ we have
\begin{eqnarray}
\frac{a(T)}{a(T^*)}=\Big(\frac{\sigma(T)}{\sigma(T^{*})}\Big)^{-1/3} \,. 
\label{aT}
\end{eqnarray}
The energy density can then be found as a function of the scale factor $\epsilon=\epsilon(a)$. The numerical results shown in Fig.~\ref{fig:ea-Tt} (left panel) demonstrate
that at a certain large $a>a_{\max}$ the energy density $\epsilon(a)$ indeed becomes negative such that the universe can never expand beyond 
$a_{\max}=a(T_{\epsilon=0})$. For $T^*=T_c$ one obtains $a_{\max}/a(T^*)\simeq 2$.
%%%%%%%%%%%%%%%%%%%%
%%%%%%%%%%%%%%%%%%%%

\begin{figure*}[!h]
\begin{minipage}{0.45\textwidth}
 \centerline{\includegraphics[width=1\textwidth]{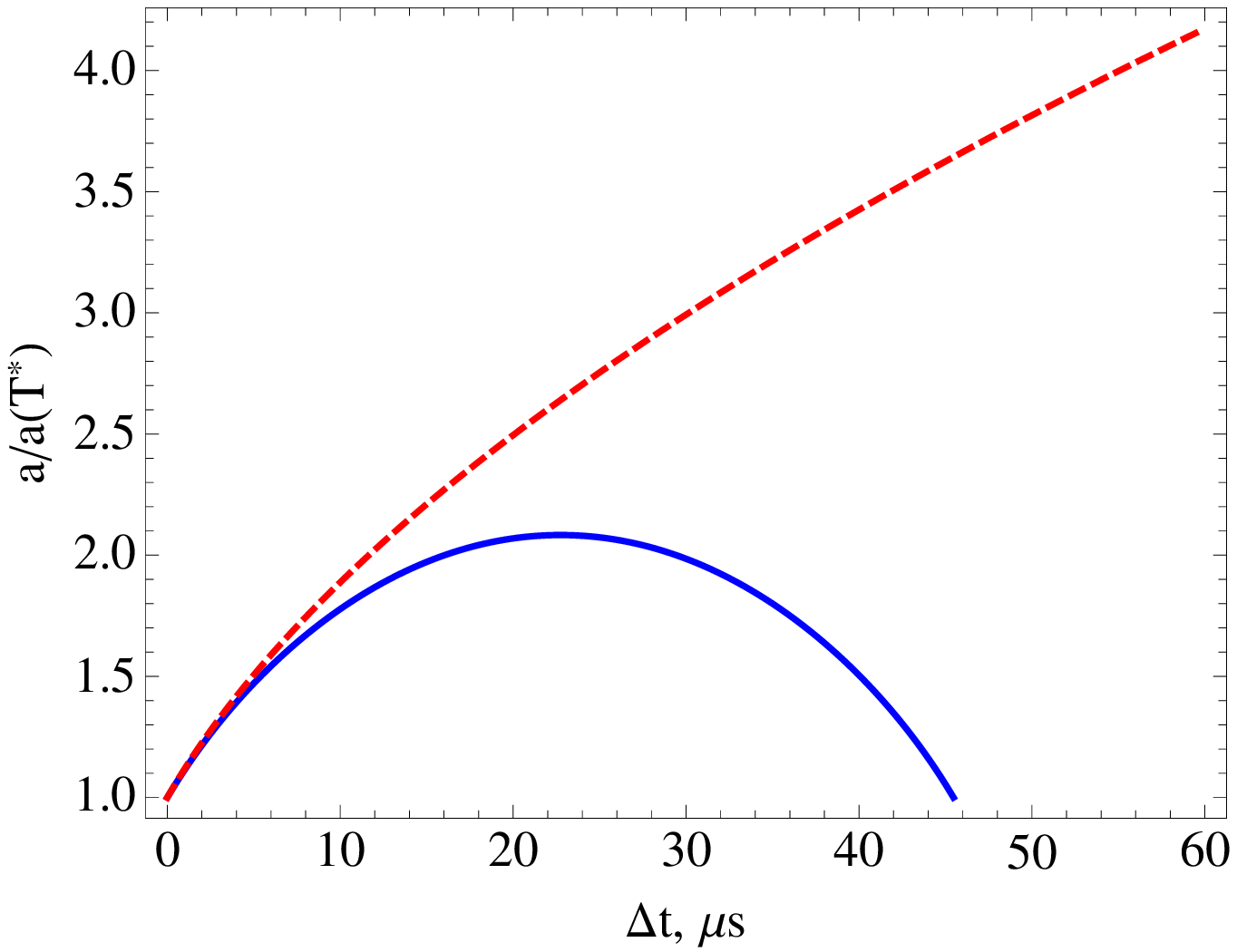}}
\end{minipage}
\begin{minipage}{0.45\textwidth}
 \centerline{\includegraphics[width=1\textwidth]{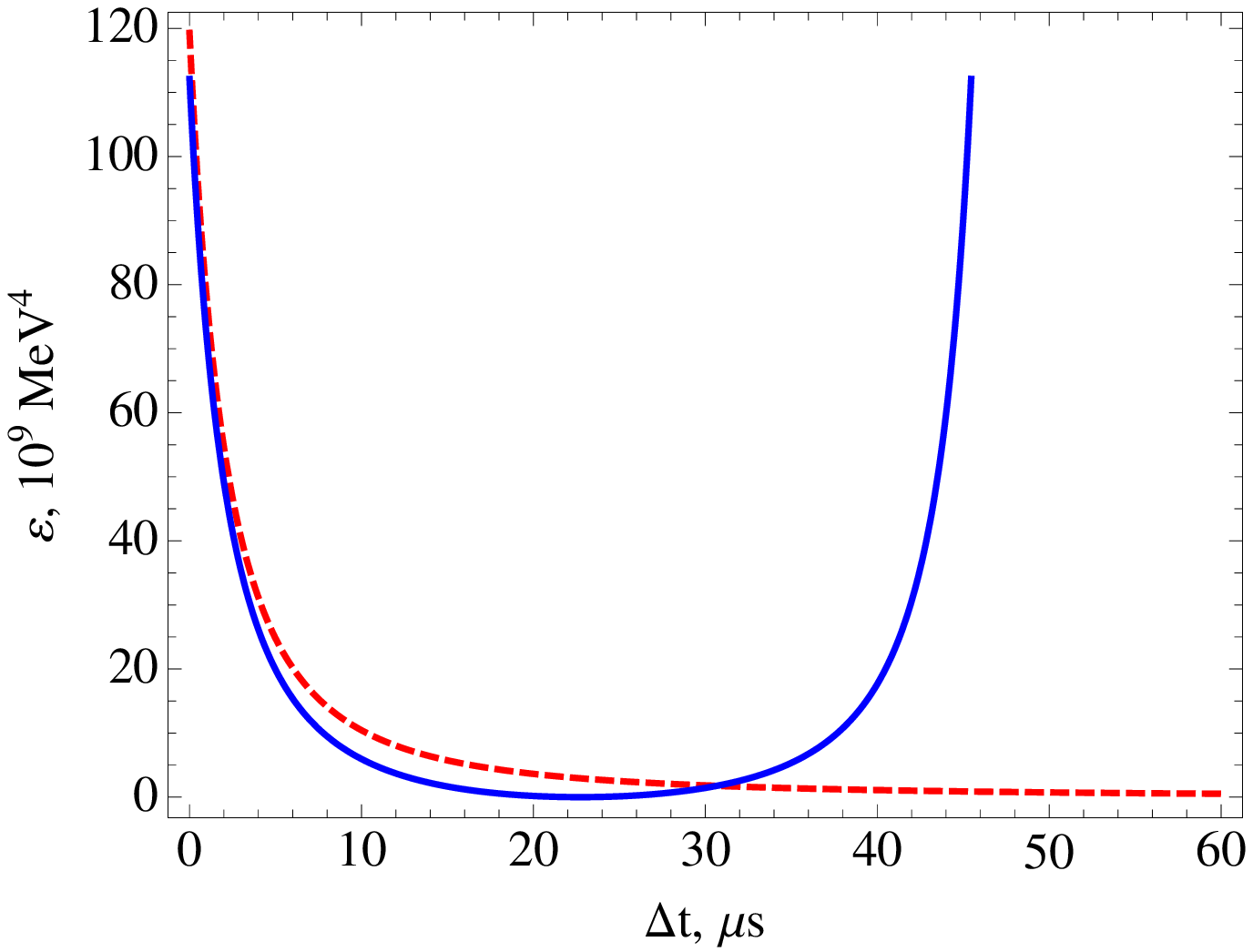}}
\end{minipage}
\caption{The scale factor as a function of time (solid line) and for the same model but with additional positive QCD-scale $\Lambda$-term (dashed line) 
is shown on the left panel. The energy density as a function of time in the meson plasma model (solid line) and 
for the same model but with additional positive QCD-scale $\Lambda$-term (dashed line) is presented on the right panel.}
\label{fig:at-et}
\end{figure*}

The time dependence of all the thermodynamical quantities can be determined from the first equation in Eq.~(\ref{Friedmann}). When energy density vanishes 
the expansion terminates and the universe begins to collapse since
\begin{eqnarray}
\frac{6}{\varkappa}\frac{\ddot{a}}{a}\Big|_{\epsilon = 0} = -3p|_{\epsilon = 0} < 0 \,,
\label{Friedmann trace1}
\end{eqnarray}
For the corresponding expansion period $\Delta t^*$, while the energy density evolves from $\epsilon(T^*)>0$ to zero, one obtains
\begin{eqnarray}
\Delta t(\epsilon)=-\frac{2}{3}\sqrt{\frac{3}{\varkappa}}\int_{\sqrt{\epsilon(T^*)}}^{\sqrt{\epsilon}}
\frac{d\sqrt{\epsilon}}{T\sigma}\,, \qquad \Delta t^*\equiv \Delta t(\epsilon=0) \simeq 22\, \mu s \,. 
\label{te}
\end{eqnarray}
For $\Delta t > \Delta t^*$, we find
\begin{eqnarray}
\Delta t(\epsilon) = \Delta t^*+\frac{2}{3}\sqrt{\frac{3}{\varkappa}}\int_{0}^{\sqrt{\epsilon}}\frac{d\sqrt{\epsilon}}{T\sigma}\,, \qquad \epsilon>0\,,
\label{te1}
\end{eqnarray}
such that functions $\epsilon(t)$, $T(t)$ and $a(t)$ can be obtained by inversion of Eqs.~(\ref{te}) and (\ref{te1}). The corresponding results are shown in 
Fig.~\ref{fig:ea-Tt} (right panel) and Fig.~\ref{fig:at-et}.
%%%%%%%%%%%%%%%%%%%
%%%%%%%%%%%%%%%%%%%

As one notices in Figs.~\ref{fig:ea-Tt} and \ref{fig:at-et} the collapse begins at $\Delta t^*$ soon after the phase transition. The collapse is followed by the warming up 
of the universe as is seen in Fig.~\ref{fig:ea-Tt}, and finally the universe reaches the temperature $T_c$ of the phase transition. The existence of the second phase transition 
during the collapse stage is an important consequence of the presence of the negative QCD vacuum contribution in cosmology. The full cycle of expansion between the two
subsequent phase transition epochs is estimated as
\begin{eqnarray}
&& \Delta t^{**}=2\Delta t^*\simeq 45 \, \mu s \,. 
\label{time cycle}
\end{eqnarray}
After the full cycle, the story may in principle repeat itself: the collapse terminates due to an unknown ``bounce'' mechanism and changes to expansion when the universe 
is filled with QGP, then it cools down and the phase transition to the meson plasma occurs with formation of the negative QCD vacuum and the universe continues to expand 
until it reaches $a_{\max}$, and so on. Apparently, such universe may oscillate in this way indefinitely long and never expands beyond $a_{\max}$ in 
the presence of the negative QCD $\Lambda$-term.

It is clear that the negative QCD vacuum contribution has to be almost exactly compensated by a positive cosmological constant for a macroscopic evolution to become possible. 
What can be the source of such a positive vacuum contribution that emerge at the QCD energy scale? So far, different approaches to the problem of negative QCD vacuum 
contribution were suggested. In Ref.~\cite{Brodsky:2009zd}, for example, it has been argued that the QCD vacuum does not have to be taken into account in cosmology at all 
as its contribution is already contained in hadron masses. Apparently, this is equivalent to the statement that the QCD vacuum has already been compensated during the
hadronisation epoch and there is no need to discuss it at any later time in the universe history. In Refs.~\cite{Pasechnik:2013sga} and \cite{Pasechnik:2016twe} it has been 
shown that under certain assumptions about a particular form of the non-perturbative gauge coupling constant it is possible to construct a generic non-perturbative solution 
for non-Abelian fields corresponding to a vacuum with a positive energy density that could compensate the ordinary QCD condensate.
%%%%%%%%%%%%%%%%%%%%%%%%%%%%%%
\begin{figure*}[!h]
\begin{minipage}{0.45\textwidth}
 \centerline{\includegraphics[width=1\textwidth]{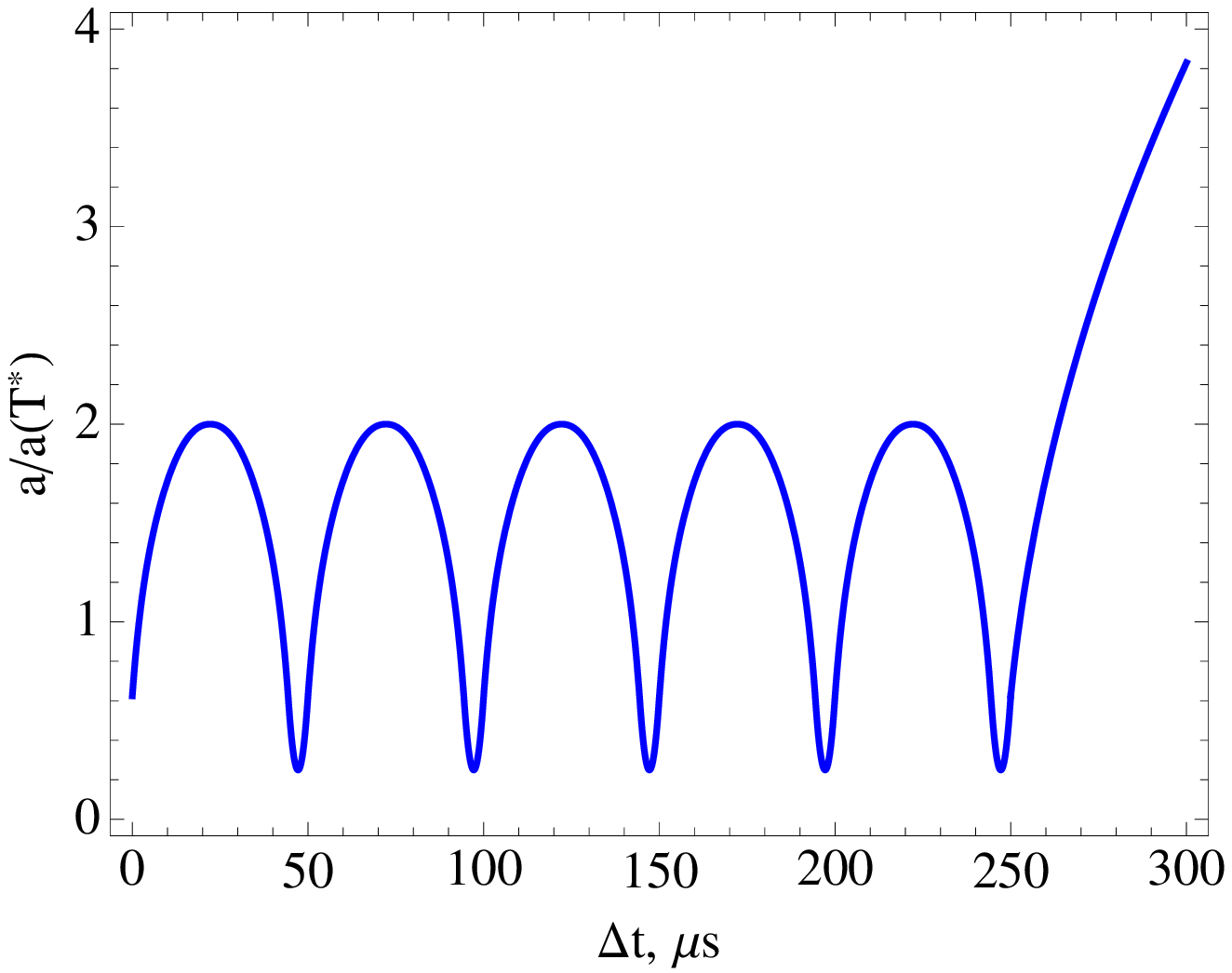}}
\end{minipage}
\begin{minipage}{0.45\textwidth}
 \centerline{\includegraphics[width=1\textwidth]{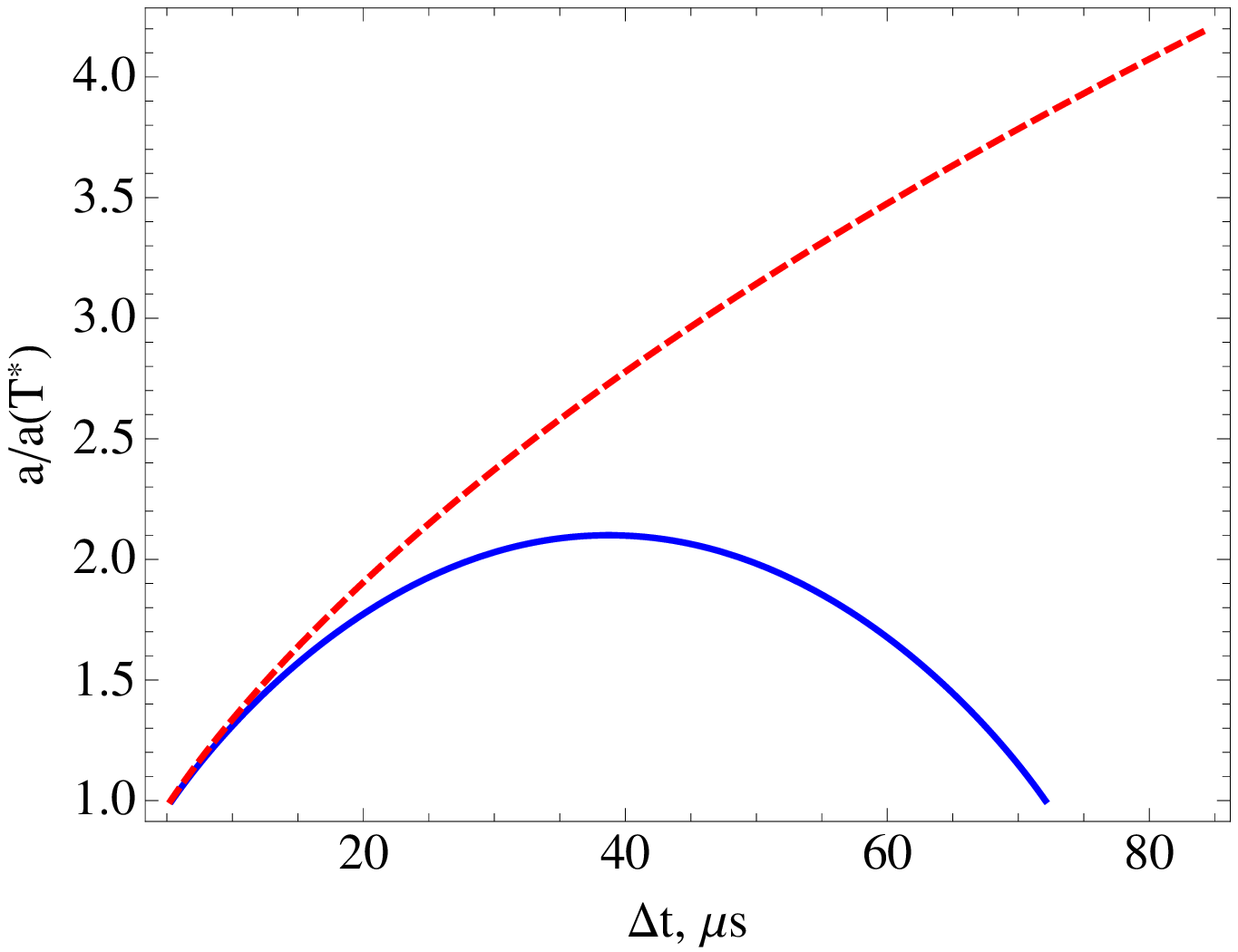}}
\end{minipage}
\caption{An illustration of the oscillating universe during QCD phase transition epoch within the hypothesis about the stochastic formation of 
the positive QCD-scale $\Lambda$-term is shown in the left panel. The results of the Bag-like model \cite{Ferroni:2008ej} for the scale factor 
as a function of time (solid line) and for the same model but with additional positive QCD-scale $\Lambda$-term (dashed line) are presented 
on the right panel.}
\label{fig:osc}
\end{figure*}
%%%%%%%%%%%%%%%%%%%%%%%%%%%%%%

While the exact compensation mechanism remains debatable, it is clear that such a compensator must exist and a positive cosmological constant at the QCD energy scale 
has to be formed at temperatures close to the QCD transition temperature together with the negative QCD term effectively eliminating the QCD vacua effects at any 
distances beyond the typically hadron length scale. According to the hypothesis about a stochastic formation mechanism of such a condensate, it may be formed 
after a series of the cycles of expansion/collapse of the universe such that after certain time scale the universe passes the phase transition scale and undergoes 
unbounded expansion as shown in Fig.~\ref{fig:osc} (left panel). This situation can be simulated by including a positive QCD-scale $\Lambda_{+}$-term by hands 
into the Friedmann equations (\ref{Friedmann}) at one of the evolution cycles without specifying a particular nature of this term such that
\begin{eqnarray}
\frac{3}{\varkappa} \frac{\dot{a}^2}{a^2}=\epsilon + \Lambda_{+} \,,\quad  
\frac{d}{dt}(\epsilon a^4)-\frac{d a}{dt}a^3(\epsilon - 3p +4\Lambda_{+}) = 0\,, \quad \Lambda_{+} + \epsilon^{\rm QCD}_{\rm vac}\simeq 0\,. 
\label{Friedmann Lambda}
\end{eqnarray}
The latter intuitive relation, although fine-tuned to many decimal digits, may not be an exact one as there should remain a small uncompensated positive cosmological constant
emerged recently in observations. The latter could be formed e.g.~in the framework of Zeldovich-Sakharov scenario as a quantum-gravity correction to the QCD vacuum density 
due to the graviton-exchange interactions in the QCD vacuum as was demonstrated in Ref.~\cite{Pasechnik:2013poa}, and other mechanisms are also possible. 
In Figs.~\ref{fig:at-et} and \ref{fig:ea-Tt} the results with $\Lambda_{+}$ obtained from Eq.~(\ref{Friedmann Lambda}) in the meson plasma model are shown 
by dashed lines. One notices that the presence of positive $\Lambda_{+}$ compensator prevents the universe from collapsing while the scenario of oscillating 
universe mentioned above is realised in the case of a very slow (stochastic) formation of $\Lambda_{+}$.

A collapsing behavior discussed above can be obtained in the framework of any effective QCD-like theory with a negative ground state density. In particular, let us consider 
the Bag-like model as an example \cite{Ferroni:2008ej} which provides a phenomenological description of both the phases and the transition in the crossover regime. 
The equation of state in the Bag-like model is defined by the following system of equations
\begin{eqnarray}
\epsilon(T) = T \frac{dp}{dT} - p \,, \qquad p = s T + B \,, \qquad  s-f(T,s)=0 \,, \label{Bag model}
\end{eqnarray}
where
\begin{eqnarray} \label{Bag model-2}
f(T,s)=c_0 \Big(\frac{T}{2\pi}\Big)^{3/2}\int_{m_{\pi}}^{\infty}d\eta \Big\{\frac{3}{4}\eta + B\frac{\eta}{4(B+sT)}
\Big\}^{3/2-\alpha}e^{\frac{\eta}{k(B+sT)^{1/4}}-\frac{\eta}{T}} \,,
\end{eqnarray}
and the model parameters are given by
\begin{eqnarray} \nonumber
\alpha = 2.3\,,\quad c_0=3642\, {\rm MeV}^{\alpha -1}\,,\quad B^{1/4}=250\, {\rm MeV}\,,\quad 
k=0.68\,,\quad m_{\pi}=139\, {\rm MeV} \,.
\end{eqnarray}
The only difference of Eqs.~(\ref{Bag model}) and (\ref{Bag model-2}) with the original formulation in Ref.~\cite{Ferroni:2008ej} is an additional vacuum term $B$ that has been added 
to the QCD energy density in order to make a shift of the vacuum density to a negative value in the hadronic phase and thus to match with the conventional QCD 
vacuum energy (in Ref.~\cite{Ferroni:2008ej} the QCD vacuum energy density is positive in the QGP phase and zero in the hadronic phase). The system (\ref{Bag model}) 
should then be solved together with the Friedmann equations (\ref{Friedmann}) or, alternatively, (\ref{Friedmann Lambda}) with a compensating positive $\Lambda$-term. 
The results of the numerical solution for the scale factor are shown in Fig.~\ref{fig:osc} (right panel). One notices that the scale factor evolves similarly to the 
one shown in Fig.~\ref{fig:at-et} demonstrating a rather universal pattern.

Another important point of the above analysis is an assumption about a mechanism of bouncing from the singularity due to which the collapse can be terminated and changed
back to expansion at some high temperatures $T>T_c$. The bouncing cosmologies is a long-debated subject in the literature, and so far various mechanisms of such a bounce
constraining the scale factor $a\neq 0$ from below were suggested. For a detailed review on this topic, see e.g.~Refs.~\cite{Novello:2008ra,Mukhanov:1991zn,
Szydlowski:2005qb,Dabrowski:1995ae} and references therein. In some of the existing scenarios, the bounce appears in ordinary gravity after the inclusion of exotic matter 
into the Einstein equations. This can be string-like or domain-wall matter or scalar fields with nonlinear oscillating potentials \cite{Dabrowski:1995ae}. In many other scenarios 
the bounce appears as a result of modifications of General Relativity by incorporating additional complicated interaction terms such as the ones in $f(R)$ gravity 
\cite{Carloni:2005ii}. The bounce can appear in ordinary gravity as well if one includes viscosity \cite{Murphy:1973zz,Novello:1980zz}. A set of other approaches is based 
on the braneworld picture \cite{Shtanov:2002mb}. 

Finally, it is worth noticing that the conclusion about the collapsing behavior and a possible series of oscillations with a consequent set of phase transitions is 
a rather model-independent statement. This statement follows just from the fact that the negative (quantum-topological) QCD vacuum found in microscopic physics 
on the general basis does participate in the universe evolution. But a particular form of the oscillations, their number and dynamics of the universe during this period 
depend on details of a complete dynamical theory of the quark-hadron plasma. The issues of dynamical formation of a positive QCD-scale vacuum contribution, 
the properties of the cosmological cycles and the picture of phase transitions during such cycles remain open for further studies.

%%%%%%%%%%%%%%%%%
\section{Conclusion}
\label{Sec:summary}
%%%%%%%%%%%%%%%%%

We have proposed a simple phenomenological model reflecting all the basic features of low-energy QCD and accounting for the lightest (pseudo)scalar meson degrees 
of freedom (the meson plasma) and the QCD vacuum in the low-temperature regime. In fact, it can be considered as an approximation of ``hadron gas'' in 
the L$\sigma$M model. One of the basic assumptions of this model is that pseudoscalar mesons do not directly interact with each other, but only via the $\sigma$-field. 
Remarkably enough, such a simplified formulation reconstructs sufficiently well the basic properties of low-energy QCD, namely, the generation of the meson masses 
by interactions with the QCD condensate, the energy density and pressure of the QCD vacuum as well as predicts the existence of the first-order phase transition. 
At the same time, it enables us to study the basic thermodynamic observables of the meson plasma fully analytically.

Of course, the model has to be improved by inclusion of heavier mesons, baryons and interaction between them which would, however, make it analytically 
unsolvable and far less transparent and can be done only in some approximations. It has been shown that the inclusion of additional heavier states into 
the model decreases the phase transition temperature. In perspective, also quarks and gluons have to be taken into account following the discussion of 
Refs.~\cite{Bowman:2010zz,Mocsy:2004ab} which would make it possible to describe the quark-gluon plasma phase above the critical temperature and, expectedly, 
has to change the phase transition to crossover.

We have extended the analysis of Ref.~\cite{Bowman:2010zz} by including the zero-point fluctuations. In comparison to other models \cite{Mocsy:2004ab,
Blaschke:2014zsa, Bowman:2010zz}, our mode incorporates additional light pseudoscalar degrees of freedom and predicts somewhat different temperature dependence
of the meson masses, namely, they decrease below $T_c$ and slowly increase above $T_c$.

It has been understood in the framework of the considered meson plasma model that the presence of the negatively-valued QCD vacuum density in cosmology leads 
to a very fast collapse of the universe shortly after the Big Bang preventing its macroscopic evolution. Adopting a realistic scenario of bouncing from the singularity, 
we have discussed the hypothesis of oscillating universe that after a series of oscillations can, in principle, pass the QCD epoch as long as a positive QCD-scale vacuum 
contribution is generated along with the negative QCD term and compensates it around the QCD transition time. If such a positive term is (stochastically) generated 
much slower than the negative one, the universe may spend a significant amount of time in the oscillating regime before the vacua compensation happens and it can turn 
to an unbounded expansion. One of the key remaining problems for further studies is formulation of a dynamical QCD-like theory where the positive QCD-scale contribution 
is naturally present.

\vspace{0.5cm}

{\bf Acknowledgments} 
The authors are indebted (with deep sorrow) to Prof. Grigory Vereshkov who has contributed to this work but passed 
away at its early stages. R.P. is partially supported by the Swedish Research Council, contract number 621-2013-428 and 
by CONICYT grant PIA ACT1406. The work has been performed in the framework of COST Action CA15213
``Theory of hot matter and relativistic heavy-ion collisions'' (THOR).

\appendix

%%%%%%%%%%%%%%%%%%%%%%%%%%%%%%%%%%%%%%%
\section{Energy-momentum tensor and free energy of the meson plasma}
\label{App:temp int}
%%%%%%%%%%%%%%%%%%%%%%%%%%%%%%%%%%%%%%%

The energy-momentum tensor of the meson plasma has the following form
\begin{eqnarray}
&&\mathcal{T}_{\mu}^{\nu}=\partial_{\mu}\sigma\partial^{\nu}\sigma +
\partial_{\mu}\pi_{\alpha}\partial^{\nu}\pi_{\alpha}+
\partial_{\mu}\eta\partial^{\nu}\eta +
\partial_{\mu}\eta'\partial^{\nu}\eta'+\partial_{\mu}\bar{K} \partial^{\nu}K +
\partial^{\nu}\bar{K} \partial_{\mu}K-\nonumber \\
&&\delta_{\mu}^{\nu}\Big( \frac{1}{2}\partial_{\lambda}\sigma\partial^{\lambda}\sigma
+2g^2 v_0^2 \sigma^2-g^4 \sigma^4 + \nonumber \\
&&\frac{1}{2}(\partial_{\lambda}\pi_{\alpha}\partial^{\lambda}\pi_{\alpha}+
\partial_{\lambda}\eta\partial^{\lambda}\eta +
\partial_{\lambda}\eta'\partial^{\lambda}\eta')+
\partial_{\lambda}\bar{K} \partial^{\lambda}K - \label{T} \\
&&\frac{1}{2}\Big[2\kappa g^2 (m_u+m_d)\sigma^2 
\pi_{\alpha} \pi_{\alpha} +\frac{2}{3}\kappa g^2 (m_u+m_d+4m_s)\sigma^2 
\eta^2 + \nonumber \\
&&\frac{4}{3}\kappa g^2 (m_u+m_d+m_s+\Lambda_{an})\sigma^2 
\eta'^2
\Big]-\kappa g^2 (m_u+m_d+2m_s)\sigma^2 \bar{K}K \Big)\,. \nonumber
\end{eqnarray}

Neglecting the chemical potential, one finds the free energy of meson plasma in the QCD condensate as the vacuum expectation value 
of the trace of the energy-momentum tensor (\ref{T}) over the spatial indices. Using Eqs.~(\ref{sigma}) and (\ref{self-cons field appr}) 
one obtains
\begin{eqnarray}
&& \mathcal{F}=\frac{1}{3}\langle\mathcal{T}_{i}^{i}\rangle =-\frac{1}{3}\big(\langle\nabla \tilde{\sigma}\nabla\tilde{\sigma}\rangle +
\langle\nabla\pi_{\alpha}\nabla\pi_{\alpha}\rangle + \langle\nabla\eta\nabla\eta\rangle + \langle\nabla\eta'\nabla\eta'\rangle +
2\langle\nabla\bar{K} \nabla K\rangle\big) -\nonumber \\
&&\Big\{ \frac{1}{2}\langle\partial_{\lambda}\tilde{\sigma}\partial^{\lambda}\tilde{\sigma}\rangle
+2v_0^2 (v^2+g^2\langle\tilde{\sigma}^2\rangle)-(v^4+6g^2v^2\langle\tilde{\sigma}^2\rangle+
3g^4\langle\tilde{\sigma}^2\rangle^2)+ \nonumber \\
&&\frac{1}{2}\big(\langle\partial_{\lambda}\pi_{\alpha}\partial^{\lambda}\pi_{\alpha}\rangle+
\langle\partial_{\lambda}\eta\partial^{\lambda}\eta\rangle +
\langle\partial_{\lambda}\eta'\partial^{\lambda}\eta'\rangle +
2\langle\partial_{\lambda}\bar{K} \partial^{\lambda}K\rangle\big) - \label{F1} \\
&&\frac{1}{2}\Big[2\kappa (m_u+m_d)(v^2+g^2\langle\tilde{\sigma}^2\rangle) 
\langle\pi_{\alpha} \pi_{\alpha}\rangle +\frac{2}{3}\kappa (m_u+m_d+4m_s)(v^2+g^2\langle\tilde{\sigma}^2\rangle)
\langle\eta^2\rangle + \nonumber \\
&&\frac{4}{3}\kappa (m_u+m_d+m_s+\Lambda_{an})(v^2+g^2\langle\tilde{\sigma}^2\rangle)
\langle\eta'^2\rangle
+2\kappa (m_u+m_d+2m_s)(v^2+g^2\langle\tilde{\sigma}^2\rangle) \langle\bar{K}K\rangle \Big]\Big\}\,. \nonumber
\end{eqnarray}
The latter expression can be simplified transformed using the basic equalities
\begin{eqnarray}
\langle\partial_{\mu}\phi\partial^{\mu}\phi\rangle=m_{\phi}^2\langle\phi^2\rangle\,,
\end{eqnarray}
and Eqs.~(\ref{eq-v}) and (\ref{e-o-m}) to a simple form
\begin{eqnarray}
&&\mathcal{F}=-\frac{1}{3}\big[\langle\nabla \tilde{\sigma}\nabla\tilde{\sigma}\rangle +
\langle\nabla\pi_{\alpha}\nabla\pi_{\alpha}\rangle+\langle\nabla\eta\nabla\eta\rangle +
\langle\nabla\eta'\nabla\eta'\rangle + 2\langle\nabla\bar{K} \nabla K\rangle\big] -\nonumber \\
&&\frac{m_{\sigma}^2}{2g^2}(\mathcal{M}^2-v^2)-2v_0^2\mathcal{M}^2+v^4+
6v^2(\mathcal{M}^2-v^2)+3(\mathcal{M}^2-v^2)^2\,, \label{F2}
\end{eqnarray}
where $\mathcal{M}$ is defined in Eq.~(\ref{eq-M}).

In calculations of the vacuum expectation values of the square field combinations appearing in Eq.~(\ref{F2}), the following 
temperature integrals emerge that are related by the recurrence relations
\begin{eqnarray}
&& J_n(T,m_{\phi})=\frac{1}{2\pi^2}\int_0^{\infty}\frac{p^{2n}dp}{\omega_{p(\phi)}}
\frac{1}{e^{\frac{\omega_{p(\phi)}}{T}}-1}\,,\quad n=0,1,2\,, \label{Jn} \\ 
&& \frac{\partial J_n(T,m_{\phi})}{\partial m_{\phi}}\Big|_T = -(2n-1)m_{\phi}J_{n-1}(T,m_{\phi})\,,  \nonumber \\
&& \frac{\partial J_n(T,m_{\phi})}{\partial T}\Big|_{m_{\phi}} = \frac{(2n-1)}{T}m^2_{\phi}J_{n-1}(T,m_{\phi})+
\frac{2n}{T}J_{n}(T,m_{\phi})\,. \nonumber
\end{eqnarray}
Besides, the contribution of vacuum oscillations
\begin{eqnarray}
J_{n\rm{(vac)}}(m_{\phi})=\frac{1}{2}\sum_{\vec{p}}\frac{p^{2(n-1)}}{\omega_{p(\phi)}} \,,\quad n=0,1,2\,, \label{Jn-vac}
\end{eqnarray}
has to be taken into account as well. They depend on the meson masses which in turn are functionals 
of the thermodynamic state of the medium. Then, the expectation values of the spatial derivatives squared in Eq.~(\ref{F2}), 
for example, can be found as follows
\begin{eqnarray}
\langle\nabla\phi\nabla\phi\rangle=J_2(T,m_{\phi})+\frac{1}{2}\sum_{\vec{p}}\frac{p^2}{\omega_{p(\phi)}}\,,
\end{eqnarray}
where
\begin{eqnarray}
\label{J2}
\omega_{p(\phi)}=\sqrt{p^2+m^2_{\phi}}\,,\qquad
J_2(T,m_{\phi})=\frac{1}{2\pi^2}\int_0^{\infty}\frac{p^4dp}
{\omega_{p(\phi)}}\frac{1}{e^{\frac{\omega_{p(\phi)}}{T}}-1}\,.
\end{eqnarray}
After regularisation, the vacuum integrals will take the following form
\begin{eqnarray}
&& J_{0\rm{(vac)}}(m_{\phi})=-\frac{1}{8\pi^2}\ln\frac{e\, m^2_{\phi}}{m^2_{\phi\rm{(vac)}}}\,, \nonumber \\
&& J_{1\rm{(vac)}}(m_{\phi})=\frac{m^2_{\phi}}{16\pi^2}\ln\frac{m^2_{\phi}}{m^2_{\phi\rm{(vac)}}}\,, \label{Jvac} \\
&& J_{2\rm{(vac)}}(m_{\phi})=-\frac{3m^4_{\phi}}{64\pi^2}\ln\frac{m^2_{\phi}}{\sqrt{e}\,m^2_{\phi\rm{(vac)}}}\,. \nonumber
\end{eqnarray}

The temperature integrals (\ref{Jn}) can be reduced to one-variable functions as follows
\begin{eqnarray}
J_n (T,m_{\phi})=T^{2 n} \widetilde{J}_n (\frac{m_{\phi}}{T})\,,\qquad 
\widetilde{J}_n (q)=\frac{1}{2\pi^2}\int_0^{\infty}\frac{z^{2n}dz}{\sqrt{z^2+q^2}(e^{\sqrt{z^2+q^2}}-1)}\,. \label{J wave}
\end{eqnarray}
where integrals $\widetilde{J}_n (q)$ behave at $q\to 0$ as
\begin{eqnarray}
&& \widetilde{J}_0(q\to 0)=\frac{1}{4\pi q}\,, \nonumber \\
&& \widetilde{J}_1(q\to 0)=\frac{1}{12}\,, \\
&& \widetilde{J}_2(q\to 0)=\frac{\pi^2}{30}\,. \nonumber
\label{zero J}
\end{eqnarray}
In the case of large $q\gg 1$, they can be expressed through the modified Bessel functions
of the second kind
\begin{eqnarray}
&& \widetilde{J}_0(q\gg 1)=\frac{1}{2\pi^2} K_{0}(q)\,, \nonumber \\
&& \widetilde{J}_1(q\gg 1)=\frac{q}{2\pi^2} K_{1}(q)\,, \\
&& \widetilde{J}_2(q\gg 1)=\frac{3 q^2}{2\pi^2} K_{2}(q)\,, \nonumber
\end{eqnarray}
and disappear exponentially at $q\to \infty$, i.e.
\begin{eqnarray}
&& \widetilde{J}_0(q\to \infty)=\frac{q^{-1/2}}{(2\pi)^{3/2}} e^{-q}\,, \nonumber \\
&& \widetilde{J}_1(q\to \infty)=\frac{q^{1/2}}{(2\pi)^{3/2}} e^{-q}\,, \\
&& \widetilde{J}_2(q\to \infty)=\frac{3 q^{3/2}}{(2\pi)^{3/2}} e^{-q}\,. \nonumber
\end{eqnarray}

%%%%%%%%%%%%%%%%%%%%%%%%
\section{Thermodynamical observables}
\label{App:obs}
%%%%%%%%%%%%%%%%%%%%%%%%

Starting from definitions in Sect.~\ref{Sec:Thermodynamics}, we obtain the expressions for the observable thermodynamical 
quantities of the meson plasma in the QCD condensate such as pressure
\begin{eqnarray}
p &=& \frac{1}{3}\Big\{J_2(T,m_{\sigma}) + 3 J_2(T,m_{\pi})+ J_2(T,m_{\eta}) \nonumber \\ 
&+& 
4 J_2(T,m_{K}) + J_2(T,m_{\eta'})\Big\} - U(v,m_{\sigma},\mathcal{M})\,, \label{pressure} 
\end{eqnarray}
the entropy density
\begin{eqnarray}
\sigma &=& \frac{1}{T}\bigg\{m_{\sigma}^2J_1(T,m_{\sigma}) + 
3m_{\pi}^2J_1(T,m_{\pi}) + m_{\eta}^2J_1(T,m_{\eta}) \nonumber \\ 
&+& 
4m_{K}^2J_1(T,m_{K}) + m_{\eta'}^2J_1(T,m_{\eta'}) + 
\frac{4}{3}\big[J_2(T,m_{\sigma}) + 3 J_2(T,m_{\pi}) \nonumber \\ 
&+&
J_2(T,m_{\eta}) + 4 J_2(T,m_{K}) + J_2(T,m_{\eta'})\big]\bigg\}\,, \label{entropy}
\end{eqnarray}
the energy density
\begin{eqnarray}
\epsilon &=& J_2(T,m_{\sigma}) + 3 J_2(T,m_{\pi}) + J_2(T,m_{\eta}) + 4 J_2(T,m_{K}) \nonumber \\ 
&+& J_2(T,m_{\eta'}) + m_{\sigma}^2J_1(T,m_{\sigma}) + 3m_{\pi}^2J_1(T,m_{\pi}) + m_{\eta}^2J_1(T,m_{\eta}) \nonumber \\ 
&+& 4m_{K}^2J_1(T,m_{K}) + m_{\eta'}^2J_1(T,m_{\eta'}) + U(v,m_{\sigma},\mathcal{M})\,, \label{energy}
\end{eqnarray}
and the heat capacity
\begin{eqnarray}
c_V &=& 3\sigma +\Big\{ 2 J_1(T,m_{\sigma}) + m_{\sigma}^2 J_0(T,m_{\sigma})
\Big\}\Big\{\frac{m_{\sigma}^2}{T}-m_{\sigma} \frac{dm_{\sigma}}{dT} \Big\}  \nonumber \\
&+& 4 \kappa \gamma (T,\mathcal{M}) \Big\{\frac{\mathcal{M}^2}{T}-\mathcal{M} \frac{d\mathcal{M}}{dT}\Big\}\,.
\label{capacity} 
\end{eqnarray}
In above expressions,
\begin{eqnarray} 
\gamma (T,\mathcal{M}) &=& \frac{3}{2} (m_u+m_d) \Big(2 J_1(T,m_{\pi}) +
m_{\pi}^2 J_0(T,m_{\pi})\Big) + \frac{1}{6}(m_u + m_d \\
&+& 4m_s) \Big(2 J_1(T,m_{\eta}) + m_{\eta}^2 J_0(T,m_{\eta})\Big)+
(m_u+m_d+2m_s) \Big(2 J_1(T,m_{K})  \nonumber \\
&+& m_{K}^2 J_0(T,m_{K})\Big) + \frac{1}{3}(m_u+m_d+m_s+\Lambda_{\rm an}) 
\Big(2 J_1(T,m_{\eta}) + m_{\eta}^2 J_0(T,m_{\eta})\Big) \,, \nonumber
\end{eqnarray}
the function $U(v,m_{\sigma},\mathcal{M})$ is defined by Eq.~(\ref{U}), and derivatives $d\mathcal{M}/dT$ and 
$dm_{\sigma}/dT$ can be obtained e.g. by differentiation of Eq.~(\ref{msM})
\begin{eqnarray}
m_{\sigma}\frac{dm_{\sigma}}{dT}&=&
\frac{2g^2}{T}\Big\{\gamma (T,\mathcal{M})+\frac{g^2}{2}\delta\Big(2 J_1(T,m_{\sigma}) + 
m_{\sigma}^2 J_0(T,m_{\sigma})\Big)\Big\}\Big(\alpha - \frac{1}{4}\delta\Big)^{-1}\,, \nonumber \\
\mathcal{M}\frac{d\mathcal{M}}{dT} &=& 
\frac{1}{4T}\Big\{\gamma (T,\mathcal{M})\Big(1-4g^4\Big[J_{0}(T,m_{\sigma}) + 
J_{0\rm{(vac)}}(m_{\sigma})\Big]\Big) + 2g^2 \Big(2 J_1(T,m_{\sigma}) \nonumber \\ 
&+&m_{\sigma}^2 J_0(T,m_{\sigma})\Big)\Big\}
\Big(\alpha - \frac{1}{4}\delta\Big)^{-1}\,, \label{dmdT}
\end{eqnarray}
and $\alpha$, $\delta$ can be found in Eqs.~(\ref{stab}), (\ref{delta}), respectively.

%%%%%%%%%%%%%%%%%

\end{document}